\documentclass[prd,12pt]{article}
\pdfoutput=1 

\usepackage{diagbox}

\usepackage{tikz}
\usepackage{amsmath,amssymb,graphicx,multirow,xspace,slashed,array,booktabs}
\usepackage[colorlinks=true,urlcolor=blue,anchorcolor=blue,citecolor=blue,filecolor=blue,linkcolor=blue,menucolor=blue,pagecolor=blue]{hyperref}
\usepackage[compress,numbers]{natbib}
\usepackage{placeins}
\usepackage{caption}
\usepackage{subcaption}
\usepackage{booktabs}
\usepackage{blindtext, rotating}
\usepackage{afterpage}
\usepackage{enumitem}
\usepackage{ marvosym }
\usepackage{authblk} 
\usepackage{verbatim}
\usepackage{soul} 
\usepackage[normalem]{ulem}
\usepackage{pifont}
\usepackage{booktabs}
\usepackage{bm}
\usepackage{cleveref}
\usepackage{siunitx}
\usepackage{mathrsfs}

\usepackage[font=footnotesize,labelfont=bf]{caption}

\usepackage{lineno}

\allowdisplaybreaks

\long\def\symbolfootnote[#1]#2{\begingroup%
\def\thefootnote{\fnsymbol{footnote}}\footnote[#1]{#2}\endgroup}

\allowdisplaybreaks

\makeatletter
	
	\@addtoreset{equation}{section}
\makeatother

\newcommand{\newc}{\newcommand}
\newc{\gsim}{\lower.7ex\hbox{$\;\stackrel{\textstyle>}{\sim}\;$}}
\newc{\lsim}{\lower.7ex\hbox{$\;\stackrel{\textstyle<}{\sim}\;$}}
\newc{\gev}{\,{\rm GeV}}
\newc{\mev}{\,{\rm MeV}}
\newc{\ev}{\,{\rm eV}}
\newc{\kev}{\,{\rm keV}}
\newc{\tev}{\,{\rm TeV}}
\newc{\MHT}{$H_T^{\text{miss}}$}
\newc{\MET}{$\slashed{E}_T$}
\newc{\MTT}{$M_{T2}$}

\def\ln{\mathop{\rm ln}}

\newc{\mz}{M_Z}
\newc{\mpl}{M_*}
\newc{\mw}{m_{\rm weak}}
\newc{\nr}[1]{N^c_R{}_{#1}}

\usepackage{accents}
\newlength{\dhatheight}

\def\beq{\begin{equation}}
\def\eeq{\end{equation}}
\newcommand{\bea}{\begin{eqnarray}\begin{aligned}}
\newcommand{\eea}{\end{aligned}\end{eqnarray}}
\def\bitem{\begin{itemize}}
\def\eitem{\end{itemize}}

\begin{document}
\baselineskip 0.6cm

\begin{titlepage}

\vspace*{-0.5cm}

\thispagestyle{empty}

\hfill MIT-CTP/5844

\begin{center}

\vskip 0.7cm

{\Large \bf
Interplay of ALP Couplings at a Muon Collider
}

\vskip 0.7cm

\vskip 0.7cm
{\large So Chigusa$^{1}$, Sudhakantha Girmohanta$^{2,3}$, \\[1ex]
Yuichiro Nakai$^{2,3}$ and Yufei Zhang$^{2,3}$}
\vskip 1.0cm
{\it
$^1$Center for Theoretical Physics, \\
Massachusetts Institute of Technology, Cambridge, MA 02139, USA, \\
$^2$Tsung-Dao Lee Institute, Shanghai Jiao Tong University, \\
No.~1 Lisuo Road, Pudong New Area, Shanghai 201210, China, \\
$^3$School of Physics and Astronomy, Shanghai Jiao Tong University, \\
800 Dongchuan Road, Shanghai 200240, China}
\vskip 1.0cm

\end{center}

\vskip 0.5cm

\begin{abstract}

Axion-like particles can couple to Standard Model gluons, electroweak gauge bosons, and massive fermions. A future multi-TeV muon collider provides a favorable environment to probe axion-like particles through multiple production channels, including vector boson fusion via electroweak gauge boson couplings and the top-associated production mediated by direct fermionic couplings. Motivated by the quality issue of the QCD axion, we focus on axion-like particles with masses and decay constants around the TeV scale. We explore how different axion-like particle couplings shape its production and decay modes, revealing a rich and intricate phenomenological landscape.

\end{abstract}

\flushbottom

\end{titlepage}


\section{Introduction}\label{intro}

Axion-like particles (ALPs) are generic pseudo-Nambu-Goldstone bosons of broken approximate global chiral $U(1)$ Peccei-Quinn (PQ) symmetry. They naturally emerge as zero modes of $p$-form gauge fields in theories with compactified extra dimensions such as string theories~\cite{Svrcek:2006yi,Arvanitaki:2009fg} or as phase components of complex scalar fields~\cite{Dine:1981rt, Kim:1979if, Shifman:1979if}. They are inspired by the QCD axion, which addresses the strong CP problem~\cite{Peccei:1977hh,Weinberg:1977ma,Wilczek:1977pj}, however, unlike the QCD axion, the mass and coupling of an ALP can vary independently. Their existence has profound implications in astrophysics, cosmology, and particle physics, driving significant experimental efforts to detect them (see, \textit{e.g.}, Refs.~\cite{ Marsh:2015xka, Beacham:2019nyx, Choi:2020rgn} for reviews).

Being representative of a compact direction, the ALP $a$ entails a mass scale $f_a$ that is referred to as its decay constant. The $U(1)$ symmetry is realized as a shift symmetry for the ALP which protects its mass. However, the ALP can obtain a mass as a result of the explicit breaking of the $U(1)$. Of particular interest are ALPs with TeV-scale masses and decay constants. Heavy axions are motivated by the axion quality problem, \textit{i.e.}, the requirement of retaining exceptionally precise global $U(1)$ symmetry, as even tiny explicit symmetry-breaking effects other than the QCD instantons (\textit{e.g.}, quantum gravity effects) could spoil the solution to the strong CP problem~\cite{Holman:1992us,Kamionkowski:1992mf,Barr:1992qq,Ghigna:1992iv,Carpenter:2009zs}. If the axion is heavy enough, this problem might be alleviated~\cite{Dimopoulos:1979pp,Holdom:1982ex,Dine:1986bg,Flynn:1987rs,Rubakov:1997vp,Berezhiani:2000gh,Fukuda:2015ana,Gherghetta:2016fhp,Dimopoulos:2016lvn,Agrawal:2017ksf,Gaillard:2018xgk,Hook:2019qoh,Gherghetta:2020keg,Gherghetta:2020ofz,Lu:2025fke}. Ref.~\cite{Lee:2021slp} envisaged embedding of an invisible QCD axion model in a warped three 3-brane geometry, where the intermediate brane is associated with the PQ breaking, while the electroweak (EW) symmetry is broken by TeV-brane localized Higgs. This simultaneously addresses the origin of the intermediate PQ-breaking scale and the EW naturalness problem. As an interesting consequence, there exist visible Kaluza-Klein excitations of the invisible QCD axion zero-mode at the TeV brane which manifests itself as a TeV scale ALP with TeV scale decay constant.~\footnote{A cosmological consequence of this model is a multi-peak gravitational wave signature corresponding to the associated phase transitions in the early Universe~\cite{Girmohanta:2023sjv}. For recent applications of multi-brane setup, also see, \textit{e.g.}, Refs.~\cite{Lee:2021wau, Girmohanta:2022giy}.}

A future multi-TeV muon collider supports a clean environment and high center of mass (CM) energy for the studies of high-energy physics, including for, \textit{e.g.}, Higgs and precision electroweak analyses~\cite{Han:2020pif, Chiesa:2021qpr, Franceschini:2021aqd} and dark matter~\cite{Han:2020uak, Asadi:2023csb, Bottaro:2021srh, Jueid:2023zxx, Belfkir:2023vpo}. Owing to their large masses, TeV scale ALPs decay promptly, making them ideal candidates for high-energy collider searches. The nature of an ALP’s couplings depends on whether the associated $U(1)$ symmetry is anomalous under QCD, EW gauge groups, or whether Standard Model (SM) fermions carry a $U(1)$ charge. As a result, ALPs can interact with gluons, EW gauge bosons, and SM fermions. Since a high-energy muon collider also effectively functions as a vector boson collider, most studies have focused on ALP interactions with EW gauge bosons~\cite{Mimasu:2014nea, Sun:2025hep, Bao:2022onq, Han:2022mzp, Inan:2022rcr, Arias-Aragon:2022iwl}, though some have also explored the ALP-top quark coupling~\cite{Chigusa:2023rrz, Inan:2025bdw, Anuar:2024qsz}. The interplay between different couplings gives rise to a diverse ALP phenomenology at a muon collider, with dominant production and decay channels varying depending on the coupling structure. This necessitates dedicated search strategies tailored to distinct phenomenological regimes, which will be the focus of this work~\footnote{For complementary studies of heavy ALPs at the LHC and HL-LHC, see Refs..~\cite{ATLAS:2022xfj, Esser:2023fdo, Bao:2022onq}.}.

We employ the ALP effective Lagrangian framework to systematically evaluate various ALP production and decay channels in a model-independent manner. We then classify the phenomenologically distinct regions of the coupling parameter space and analyze them separately. The key distinctions arise from whether the ALP is produced via vector boson fusion or in association with a top-quark, and whether it decays into a pair of gauge bosons or a top-quark pair. We demonstrate that a dedicated forward muon (FM) detector is essential for isolating the ALPs produced by vector boson fusion (VBF) while effectively suppressing a wide range of non-VBF SM backgrounds. For the parameter region where the ALP decays into a pair of top quarks, the choice of the cone size parameter $R$ in jet clustering, optimized based on the ALP mass range, as well as correlations in the azimuth angles and transverse momenta of the jets, further refine the signal selection. We reconstruct the peak structure in the dijet invariant mass spectrum and apply a likelihood analysis to quantify the signal significance. Finally, we project the resulting sensitivity of a future muon collider to the ALP parameter space. We also discuss the challenges associated with reconstructing ALP decays into gluons, while a strategy to make further progress is outlined.

The remainder of the paper is structured as follows. Section~\ref{sec:setup} introduces the ALP effective theory framework, outlining the various production and decay channels and categorizing them based on distinct phenomenological regions. Sections~\ref{sec:VT},~\ref{sec:TT},~\ref{sec:VGTG} provide a detailed analysis of each region, including the event selection strategy and statistical treatment. Section~\ref{conclusion} contains our conclusions and further discussions.

\section{Setup}
\label{sec:setup}
Here we define our effective theory of ALP and introduce its couplings to the SM. Various ALP production and decay channels are analyzed, identifying phenomenologically distinct regions in the coupling parameter space. Among these, two regions are considered in the literature, while three new regions are yet to be explored. 

\subsection{EFT Framework}\label{setup}
Let us extend the SM with a pseudoscalar ALP $a$ with mass $m_a$ and decay constant $f_a$. As we are interested in the dynamics of TeV-scale ALPs, the effective Lagrangian should be full SM gauge group invariant, \textit{i.e.}, invariant under $G_{\rm SM} = SU(3)_c \otimes SU(2)_L \otimes U(1)_Y$~\cite{Bauer:2017ris, Bauer:2020jbp, Brivio:2017ije},
\begin{multline}
    {\cal L}_{\rm eff} = {\cal L}_{\rm SM} + \frac{1}{2} (\partial_\mu a) (\partial^\mu a) - \frac{1}{2} m_a^2 a^2 - \left( \frac{g_s}{4 \pi}\right)^2 C_{{G}} {\cal A}_{\widetilde{G}} -  \left( \frac{g}{4 \pi}\right)^2 C_{{W}} {\cal A}_{\widetilde{W}} \\ - \left( \frac{g'}{4 \pi}\right)^2 C_{{B}} {\cal A}_{\widetilde{B}} - \frac{C_{a \Phi}}{2} \frac{\partial_\mu a}{f_a} \sum_{\psi=Q, L} \bar \psi \gamma^\mu \gamma_5 \sigma_3 \psi + \text{h.c.} \ ,
    \label{Eq:ALP_EFT}
\end{multline}
where the $Q, L$ represent the SM quark and lepton doublets, and $\sigma_3$ acts on the weak isospin space. $g_s$, $g$, $g'$ are the corresponding coupling constants of $SU(3)_c$, $SU(2)_L$, and $U(1)_Y$, respectively, while the symbol $A_{\widetilde{X}}$ denotes
	\begin{equation}
		{\cal A}_{\widetilde{X}} = X_{\mu \nu} \widetilde{X}^{\mu \nu} \frac{a}{f_a} \ ,
	\end{equation}
for a gauge field $X \in \{G, W, B\}$, where $X^{\mu \nu}$ and $\tilde{X}^{\mu\nu}$ are the field strength tensor and its dual $\widetilde{X}^{\mu \nu} = \epsilon^{\mu \nu \rho \sigma} X_{\rho \sigma}/2$, respectively. In Eq.~\eqref{Eq:ALP_EFT}, higher-order powers of $a/f_a$ have been neglected, and the coefficients $C_{W}, C_{B}, C_{G}, C_{a\Phi}$ represent the corresponding coupling strengths. We have followed the convention where the loop factor is not absorbed in the definition of the ALP-gauge boson couplings. After the electroweak symmetry breaking, the ALP couplings with the $B_\mu$ and $W_\mu^3$ get converted into the ALP couplings with the $\gamma$ and $Z$ boson. 

Utilizing Eq.~\eqref{Eq:ALP_EFT}, one can evaluate the partial decay widths of the ALP to dominant two-body SM final states as follows~\cite{Aloni:2018vki, Bao:2022onq}
\begin{align}
\label{Eq:partial_widths}
     \Gamma(a \to f \bar f)  &= \frac{3 m_a m_f^2}{8 \pi f_a^2} \left| C_{a\Phi} \right|^2 \sqrt{1- \frac{4 m_f^2}{m_a^2}}\ , \\
     \Gamma(a \to \gamma \gamma)  &= \frac{\alpha_{\rm em}^2 m_a^3}{64 \pi^3 f_a^2} \left( C_{B} + C_{W} \right)^2\ , \\ 
     \Gamma(a \to g g)  & = \frac{\alpha_{s}^2(m_a) m_a^3}{8 \pi^3 f_a^2}  \left| C_{G} \right|^2 \left[ 1 + \frac{83}{4} \frac{\alpha_s (m_a)}{\pi} \right]\ , \\ 
     \Gamma(a \to \gamma Z)  & = \frac{\alpha_{\rm em}^2 m_a^3}{32 \pi^3 f_a^2} \left( C_W \cot \theta_W - C_B \tan \theta_W \right)^2 \left[ 1- \left(\frac{m_Z}{m_a}\right)^2\right]^3\ , \\ 
    \Gamma(a \to Z Z)  & = \frac{\alpha_{\rm em}^2 m_a^3}{64 \pi^3 f_a^2} \left( C_W \cot^2 \theta_W + C_B \tan^2 \theta_W \right)^2 \left[ 1- \left(\frac{2 m_Z}{m_a}\right)^2\right]^{3/2}\ , \\ 
     \Gamma(a \to W^+ W^-)  & = \frac{\alpha_{\rm em}^2 m_a^3}{32 \sin \theta_W^4 \pi^3 f_a^2} \left| C_W \right|^2  \left[1-\left(\frac{2m_W}{m_a}\right)^2 \right]^{3/2} \ ,
\end{align}
where $f$ is a massive SM fermion with mass $m_f$, $\alpha_{\rm em}$ is the fine structure constant, $\alpha_s(\mu) \equiv g_s^2(\mu)/(4\pi)$ is the running strong coupling constant in $\overline{\rm MS}$ scheme, $\theta_W$ is the weak mixing angle and we have included the one-loop correction to the $a \to g g$ decay rate. The running coupling constants are to be evaluated at the scale $m_a$.

\subsection{ALP production and decay channels}
\label{sec:regions}

\begin{figure}[t]
    \centering
    \includegraphics[width=0.75\textwidth]{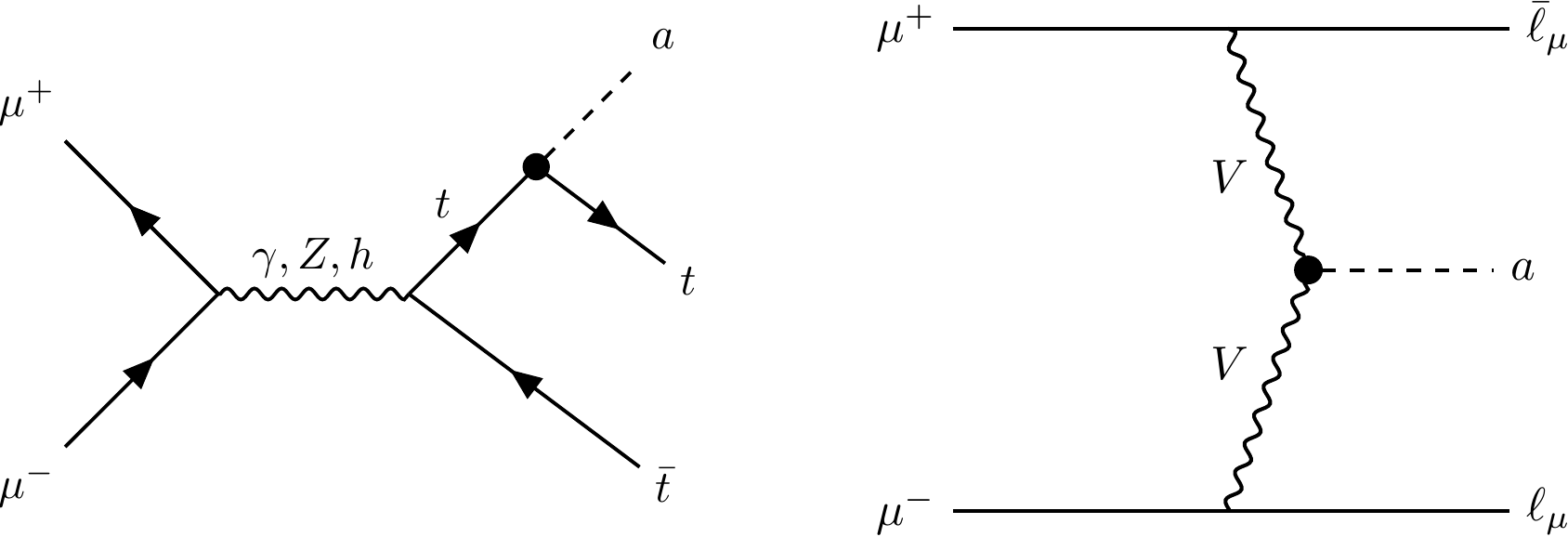}
    \caption{Representative Feynman diagrams for the top-associated production (left) and VBF of ALP. Here $V=\gamma, Z, W$ and $\ell_\mu = \mu, \nu_\mu$ depending on $V$.}
    \label{fig:VBF_top}
\end{figure}

In Fig.~\ref{fig:VBF_top}, we show a few representative diagrams from a class of ALP production diagrams, classified broadly into the top-associated production channel (left) or the VBF production channel (right). The top quark is the most relevant for TeV scale ALPs as the corresponding coupling scales with the SM quark/lepton mass. The branching ratio of the ALP decay to any final state (f.s.) can be evaluated as
\begin{equation}
    {\cal B} (a \to {\rm f.s.}) = \frac{\Gamma(a \to {\rm f.s.})}{\Gamma_{\rm total}} \ ,
\end{equation}
where $\Gamma(a \to {\rm f.s.})$ denotes the partial decay width of the specific channel $a \to {\rm f.s.}$, while $\Gamma_{\rm  total}$ denotes the total ALP decay width. Depending on the relative size of the ALP-SM couplings, the dominant decay channel of ALP will be different. 
\begin{figure}[t]
    \centering
    \includegraphics[width=\textwidth]{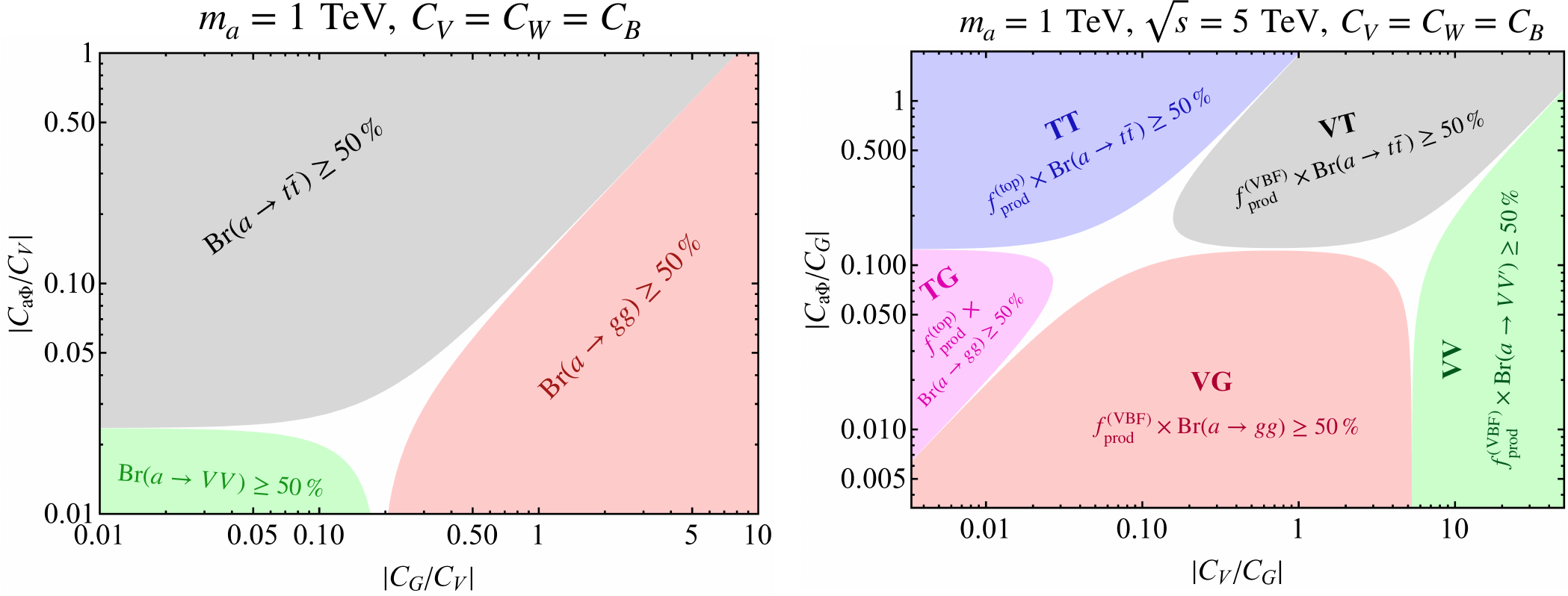}
    \caption{Interplay of ALP couplings in determining the dominant production and decay channels of the ALP.}
    \label{fig:ALP_production_decay}
\end{figure}
For illustrative purposes, in Fig.~\ref{fig:ALP_production_decay}, we demonstrate the interplay of ALP-SM couplings in determining the dominant production and decay channels for a $1$ TeV ALP in a future muon collider. In the left subfigure, we show the different dominant ALP decay channels in the ALP-fermion ($C_{a\Phi}$) and ALP-gluon coupling ($C_G$) parameter subspace, normalized with the choice $C_V=C_W=C_B$. The ALP dominantly decays to $t \bar t$ in the gray region, to $gg$ in the red region, to electroweak gauge bosons $VV'$ with $V, V'=\gamma, Z, W$ in the green region, while the ALP has multiple competitive decay channels in the marginal white region. Similarly, in the right sub-figure, we show the dominant,  phenomenologically distinct regions for the ALP production and decay channels. Specifically, we identify the regions for which $f_{\rm prod}^{(\rm ch)} \times {\cal B}(a \to {\rm f.s.})>50\%$, where $f_{\rm prod}^{(\rm ch)}$ is the fraction of the ALP production cross-section in a particular channel `$\rm ch$'. The dominant production channel can be either the top-associated production $\mu^+ \mu^- \to a t \bar t$, or the VBF $\mu^+ \mu^- \to \ell_\mu \bar \ell_\mu a$, where $\ell_\mu = \mu, \nu_\mu$ as in Fig.~\ref{fig:VBF_top}. In the right subfigure, we fix $C_V=C_{W}=C_{B}$, and depict the distinct regions in the ALP-fermion ($C_{a\Phi}$) versus ALP-EW coupling ($C_V$) subspace, normalized to $C_G$. Here we have assumed $C_G$ is non-vanishing, otherwise, one can plot the individual couplings to obtain similar behavior. To estimate  $f_{\rm prod}^{(\rm ch)}$, we utilized MadGraph5$\_$aMC~\cite{Alwall:2014hca} to evaluate the production cross-section for a particular coupling and the fact that it scales quadratically with the coupling involved. In particular, we identify five phenomenologically distinct regions and label them as follows~\footnote{Here we have ignored the subdominant associated vector boson production channel $Va$ (see Ref.~\cite{Bao:2022onq}), however, it can be potentially important for ALP mass near the threshold $m_a \sim \sqrt{s}$.}. 
\begin{itemize}
    \item \textbf{Region TT}: The ALP production dominantly proceeds through the top-associated production and the ALP decays to $t \bar t$. This corresponds to the 4-top channel $\mu^+ \mu^- \to t \bar t (a \to t \bar t)$.

    \item \textbf{Region TG}: The dominant ALP production is the top-associated production, and the decay channel is $a \to gg$, corresponding to the process $ \mu^+ \mu^- \to t \bar t (a \to gg)$. 

    \item \textbf{Region VT}: The ALP production happens through VBF dominantly, while it decays to $t \bar t$, \textit{i.e.}, the channel is $\mu^+ \mu^- \to \ell_\mu \bar \ell_\mu (a \to t \bar t)$, where $\ell_\mu = \mu, \nu_\mu$ for charged or neutral boson fusion, respectively.

    \item \textbf{Region VG}: ALP is produced dominantly through VBF and decays to $gg$, \textit{i.e.}, the process is $\mu^+ \mu^- \to \ell_\mu \bar \ell_\mu (a \to g g)$, with $\ell_\mu = \mu, \nu_\mu$.

    \item \textbf{Region VV}: VBF is the dominant production channel for ALP, while it also decays to EW gauge bosons, including $a \to W^+ W^-, \gamma Z, Z Z, \gamma \gamma$.
\end{itemize}
The sixth possible region, namely TV, where the production is through the top-associated channel but the ALP decays to electroweak gauge bosons never dominate because if the ALP-top coupling is substantial enough to contribute to the production dominantly, then the ALP decay to tops will dominate over the decays to electroweak gauge bosons due to the large top mass and the loop-suppressed ALP-gauge boson coupling. Finally, there could be a possibility that the ALP dominantly decays to some dark sector particles, but also couples considerably to the SM such that it can be produced in a collider. We will not entertain this possibility further as we focus on the on-shell production of the ALP such that its decay products can be reconstructed and the ALP mass can be inferred from the invariant mass of its decay products.

Motivated by the fact that a high-energy muon collider also acts as an electroweak vector boson collider, considerable attention has been given to the region VV~\cite{Bao:2022onq, Han:2021udl}, where the VBF produced ALP decays to electroweak gauge bosons. It has been pointed out that the diphoton decay channel is relatively clean and the diphoton invariant mass can be used to reconstruct the ALP event. The present authors, in Ref.~\cite{Chigusa:2023rrz} analyzed the region TT, which utilized the ALP-top coupling solely and demonstrated that utilizing the hadronic decay channels of the top quark, a dijet mass resonance search can be used to probe this TT region for an on-shell ALP, thanks to the substantial boost factor provided by the high-energy fundamental muon beams. However, a full mapping of the phenomenologically distinct regions was not presented in the past, and here we identify three new regions, namely TG, VT, and VG.

\section{Region VT}
\label{sec:VT}

{In the VT region corresponding to the gray parameter space in Fig.~\ref{fig:ALP_production_decay} (right), as defined in sec.~\ref{sec:regions}, the dominant ALP production is through the VBF process, while the ALP decays to $t \bar{t}$ predominantly. To illustrate, for $m_a = 1$ TeV, $|C_{a\Phi}|/f_a = 6$ TeV$^{-1}$, $|C_{W}|/f_a = |C_{B}|/f_a = 10$ TeV$^{-1}$, and $\sqrt{s}=5$ TeV (see Fig.~\ref{fig:ALP_production_decay}), the ALP branching ratio is ${\cal B} (a \to t \bar t) \gtrsim 99.5\%$, which is almost insensitive to the choice of $C_{G}$, as long as we remain inside the VT region. Further, for this illustrative set of parameters, $\Gamma (a \to t \bar t) \simeq 1.2 \times 10^2$ GeV, as obtained from MadGraph5$\_$aMC~\cite{Alwall:2014hca} simulation, which agrees with Eq.~\eqref{Eq:partial_widths}. Hence, a typical ALP flight length is $L_{a}  = c \tau_a \beta_a \gamma_a \simeq 10^{-2}$ fm, where $c$ denotes the speed of light, $\tau_a$ is the ALP lifetime, and $\beta_a \gamma_a \simeq \sqrt{s}/m_a$ is the maximum possible combination of the velocity and boost factor for the ALP. Therefore, the ALP decays promptly to $t \bar t$ pair, which can be utilized to reconstruct the ALP through the dijet invariant mass. We will use the hadronic decay products of the top to reconstruct the boosted top jets and identify them with the jet mass reconstructed as such. This is a preferred method of top reconstruction in a high-energy muon collider, thanks to the fundamental nature of the colliding particles, which imparts a significant boost to the top quark produced such that its hadronic decay products lie in a single jet cone. This situation is to be contrasted with a hadronic collider as the available collision energy is shared among the parton level fundamental constituents~\cite{ATLAS:2021kqb, ATLAS:2023ajo, CMS:2023ftu}. Having established the reconstruction strategy for the ALP-generated signal event, and its production channel, let us now turn to possible SM backgrounds.} 

SM processes with multiple top-jets in the final state, including both VBF and non-VBF production channels should be considered as backgrounds, while events with additional non-top-jets, which can be misidentified as tops may also contribute significantly to the background. Nevertheless, as the VBF processes are enhanced in the soft and collinear regions for the vector boson involved, they are often accompanied by outgoing leptons primarily in the forward region, \textit{i.e.}, the region with high pseudorapidity ($\eta$). In our analysis, we use the fact that a plethora of non-VBF SM backgrounds can be suppressed if the forward muons (FMs) can be tagged with an FM detector.

\begin{table*}[!t]\centering
	\setlength{\tabcolsep}{3mm}
    \resizebox{\textwidth}{!}{ 
	\begin{tabular}{|c||c|c|c|c|c|} \hline
		&$ \ell_\mu \bar \ell_\mu a  \left(1 \, {\rm TeV}\right)$ & $ \ell_\mu \bar \ell_\mu a \left(1.5 \, {\rm TeV}\right)$ &  $\ell_\mu \bar \ell_\mu a \left(2.5 \, {\rm TeV}\right)$ & $ \ell_\mu \bar \ell_\mu t\bar{t}$ & $ \ell_\mu \bar \ell_\mu W^+ W^-$
		\\ \hline
		$\sigma$[pb] & 0.002475 & 0.001995 & 0.001295 &  0.0096 &  1.356 \\ \hline
		
	\end{tabular}
    }
 \vspace{0.1cm}
	\caption{
	Cross-sections for the signal and principal backgrounds {in the VT region} for the chosen benchmark parameters, $\{|C_{W}|,|C_{B}|, |C_{G}|,|C_{a \Phi}|\}/f_a=\{10, 10, 0, 6\} \ {\rm TeV}^{-1}$ at $\sqrt{s} = 10  \ {\rm TeV}$ {for three different ALP masses as mentioned in the parenthesis}.}
 \label{tab:VT_cross_sections}
\end{table*}

The importance of the FM detectors has been motivated for high-energy muon colliders~\cite{Accettura:2023ked, AlAli:2021let}, as they may bring significant advantages, including for, \textit{e.g.}, extracting the SM $hZZ$ couplings by isolating the $ZZ$-fusion production of the Higgs boson~\cite{Li:2024joa}. To suppress the beam-induced backgrounds (BIBs) as a result of unstable muon decays, two tungsten cone-shaped nozzles are designed to be placed around the beampipe, which typically restricts the angular coverage of the detection region to $|\eta| \lesssim 2.5$, corresponding to $10^{\circ} \lesssim \theta \lesssim 170^{\circ}$~\cite{MuonCollider:2022ded}. Since the TeV-scale FMs can penetrate these shields, the FM detectors can be placed beyond them, covering the region with $|\eta| \gtrsim 2.5$.

We assume that muons with $2.5 < |\eta| < 8$ can be tagged as FMs~\cite{Li:2024joa}. We find that the non-VBF SM backgrounds are reduced as such by requiring tagged FMs. Note that the $WW$-fusion produces outgoing neutrinos, which can not be tagged, and the requirement of the tagged FMs essentially selects fewer signal events, only associated with the neutral VBF channels. Still, the reduction of the plethora of non-VBF SM backgrounds makes it an invaluable filter. As an illustration, in Table~\ref{cutflow_table_5TeV}, we show the raw number of events for one of the principal backgrounds in the VT region, namely $\mu^+ \mu^- \to \ell_\mu \bar \ell_\mu t \bar t$, which includes events from both the VBF and non-VBF processes. However, non-VBF induced $\mu^+ \mu^- \to \mu^+ \mu^- t \bar t$ events have outgoing muons that are distributed in the whole solid angle, unlike the VBF induced events, where the outgoing muons are primarily in the forward region. Therefore, demanding at least one FM for selecting a candidate event significantly trims down the non-VBF part of this background, as well as the $WW$-fusion events. Among the remnant SM VBF backgrounds, $\mu^+ \mu^- \to \mu^+\mu^-W^+ W^-$ is the dominant one. We will show that requiring two top jet candidates will cut down this background significantly. {The other subdominant backgrounds are SM VBF $t\bar{t}h$ ($\sigma = 4.2 \times 10^{-5} \ {\rm pb}$) and $t\bar{t}z$ ($\sigma = 8.0 \times 10^{-5} \ {\rm pb}$). Their cross-sections are two orders smaller than the signal cross-section and we will ignore them in the following analysis. Furthermore, these are removable almost entirely by requiring the FM cut. 

Let us now consider a specific choice of parameter set for the sake of illustration for the VT region, namely $\{|C_{W}|, |C_{B}|, |C_{G}|, |C_{a \Phi}|\}/f_a=\{10, 10, 0, 6\}{\ \rm TeV}^{-1}$. In Table~\ref{tab:VT_cross_sections}, we show the production cross-sections for the signal and background processes with CM energy $\sqrt{s}=10\,\mathrm{TeV}$ {for the benchmark parameter choice mentioned above, for three different ALP masses} $m_a = \{1,1.5,2.5\} {\ \rm TeV}$. Here, $C_W = C_B$ is assumed for simplicity, while analysis for a different choice is straightforward. Since the signal analysis, which will be described later, is conducted in three distinct ALP mass regions to optimize the signal selection efficiency, we consider three corresponding ALP masses in this study.

We now comment on the BIB mitigation techniques and their consequences that are relevant to all parameter regions discussed in this paper.
A big challenge we need to tackle is to distinguish true jets from fake jets due to the BIB.
Utilizing the specific characteristics of the fake jets associated with BIBs, such as their low transverse momenta $p_{\rm T}$, asynchronous time of arrival, displaced origin, and high $|\eta|$ can also help mitigate the fake jet-induced backgrounds~\cite{MuonCollider:2022ded}.}
Following the filtering procedure of jets introduced in ref.~\cite{MuonCollider:2022ded}, it is possible to reduce the effect of the BIBs to a negligible level, while keeping $\sim 90\,\%$ of true jets intact.
In our analysis, we randomly pick up the $90\,\%$ of jets to simulate this jet reconstruction efficiency and neglect the BIB effect.\footnote{
This procedure is based on an approximation that the filtering procedure does not significantly modify the distribution of jets in their phase space.
It is beyond the scope of this paper to perform a more detailed analysis of this procedure.
}

\subsection{Event generation and cut-flow}
\label{sec:eventgen_VT}

The patron level events are generated by using the MadGraph5$\_$aMC~\cite{Alwall:2014hca} and then they are processed by Pythia8~\cite{Bierlich:2022pfr} for matching of jets, showering, and hadronization. The jet clustering is performed by FastJet~\cite{Cacciari:2011ma}. Detector simulation is performed by Delphes~\cite{deFavereau:2013fsa} using the muon collider detector card~\cite{delphesTalk}. We implement the ALP model Lagrangian in FeynRules~\cite{Alloul:2013bka} following Ref.~\cite{Brivio:2017ije} and then generate a model file compatible with MadGraph5$\_$aMC~\cite{Alwall:2014hca}. To avoid infrared (IR) singularities, a parton-level pre-selection is applied when generating both signal and background samples~\cite{PhysRevD.109.073009}:
\begin{equation}
		p_T(l,j) > 5 {\ \rm GeV}, \ \ \ p_T(\gamma) > 1 {\ \rm GeV}, \ \ \  0 < \eta(l) < 10, \ \ \  \Delta R(jj,ll) > 0.2 ,
\end{equation}
where the $l$, $j$ are leptons and jets, respectively, {$p_{T}$ denotes the transverse momentum, and $\eta$ represents pseudo-rapidity, while $\Delta R (j_1 j_2) = \sqrt{(\eta(j_1)-\eta(j_2))^2+(\phi(j_1)-\phi(j_2))^2}$ is the angular distance between two jets $j_1$, $j_2$, and similarly for the leptons.} As the TeV-scale ALP in the VT region dominantly decays to a pair of top quarks, improving the top reconstruction efficiency is desired. Proper reconstruction methods depend on the momenta of the top pair, but looking for boosted top jets focusing on the hadronic decay of the top is a promising way in the high energy environment of a muon collider. The radius of top jets can be large depending on the size of the boost. Therefore, choosing a proper cone size $R$ for the jet reconstruction is necessary to collect the top jets from ALP. To maximize the signal events with two potential top candidates, we choose $R=1.0$ and $R=0.8$ for $\sqrt{s}=5$ TeV and $\sqrt{s}=10$ TeV, respectively, where we have used the $k_T$ jet-clustering algorithm.

\begin{figure}[!t]
	\centering
		\begin{subfigure}{0.45\textwidth}
		\centering
		\includegraphics[width=\textwidth]{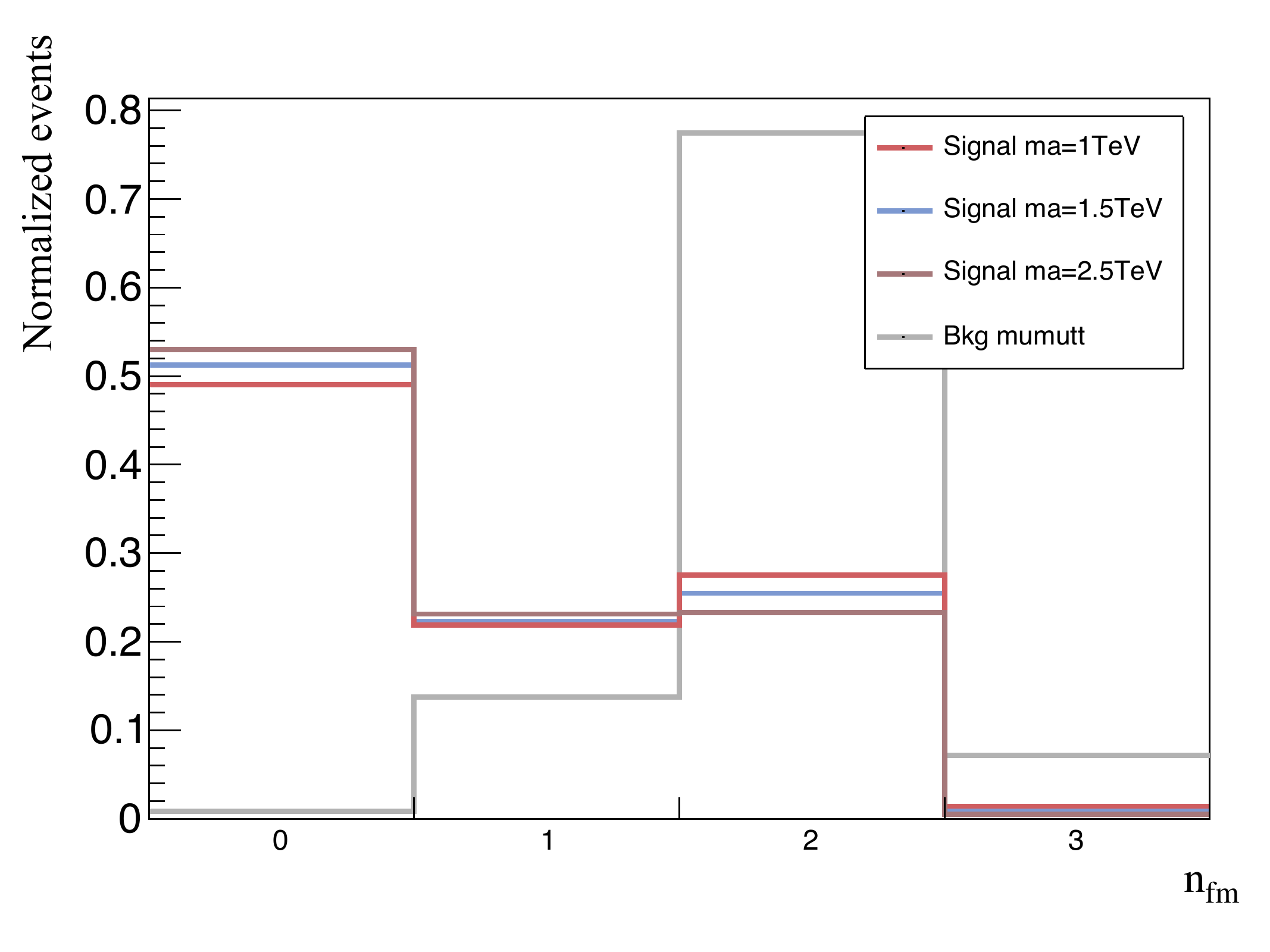}
		\subcaption{Number of FMs}
		\label{fig:fm}
		\end{subfigure}   
\hspace{0.8cm}
		\begin{subfigure}{0.45\textwidth}
		\centering
		\includegraphics[width=\textwidth]{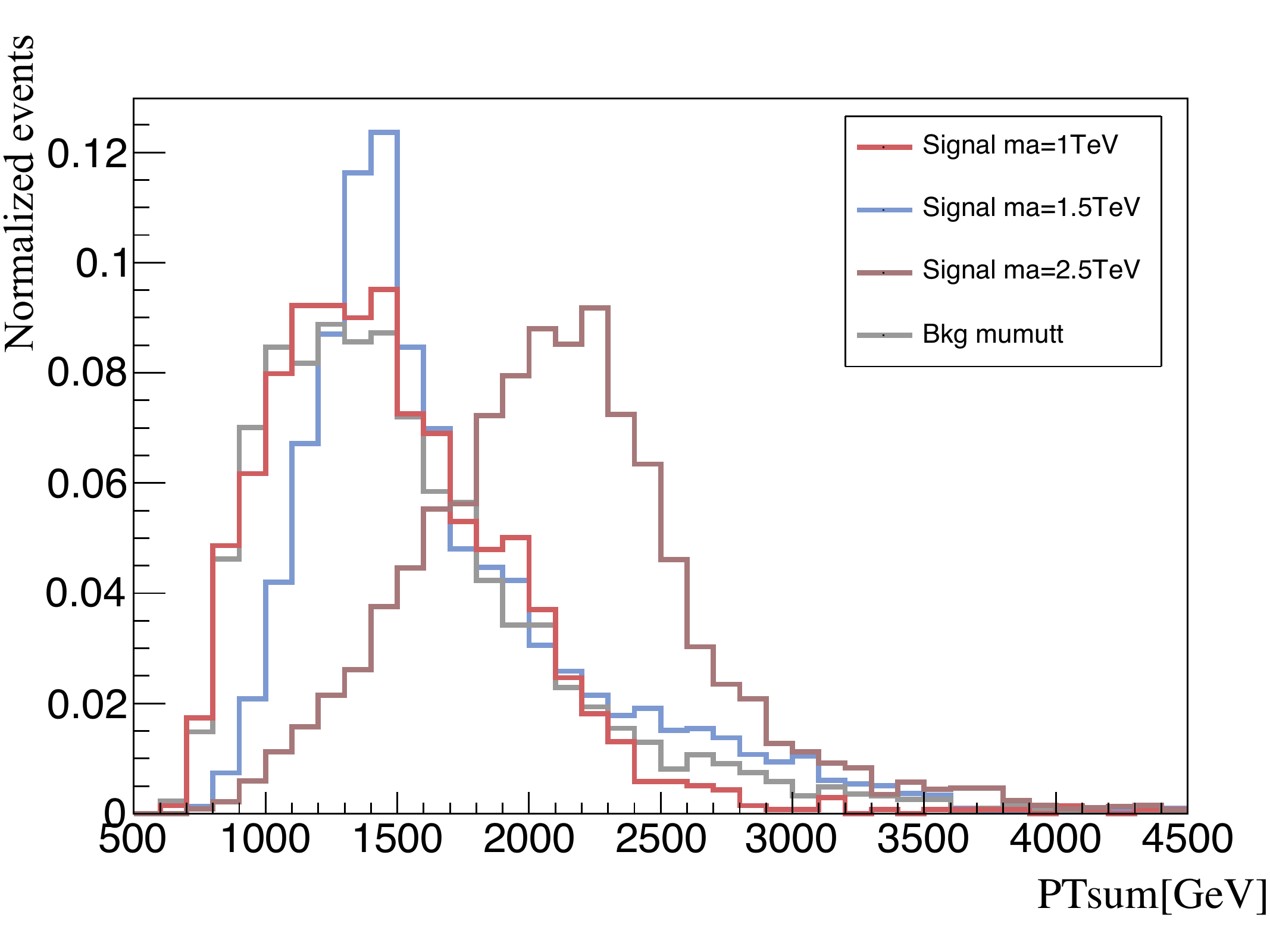}
		\subcaption{$\mathrm{PT}_{\mathrm{sum}} \ ({\rm n}_{\rm fm}=1)$}
		\label{fig:PT_sum1}
		\end{subfigure}
\hspace{0.8cm}
 	\begin{subfigure}{0.45\textwidth}
		\centering
		\includegraphics[width=\textwidth]{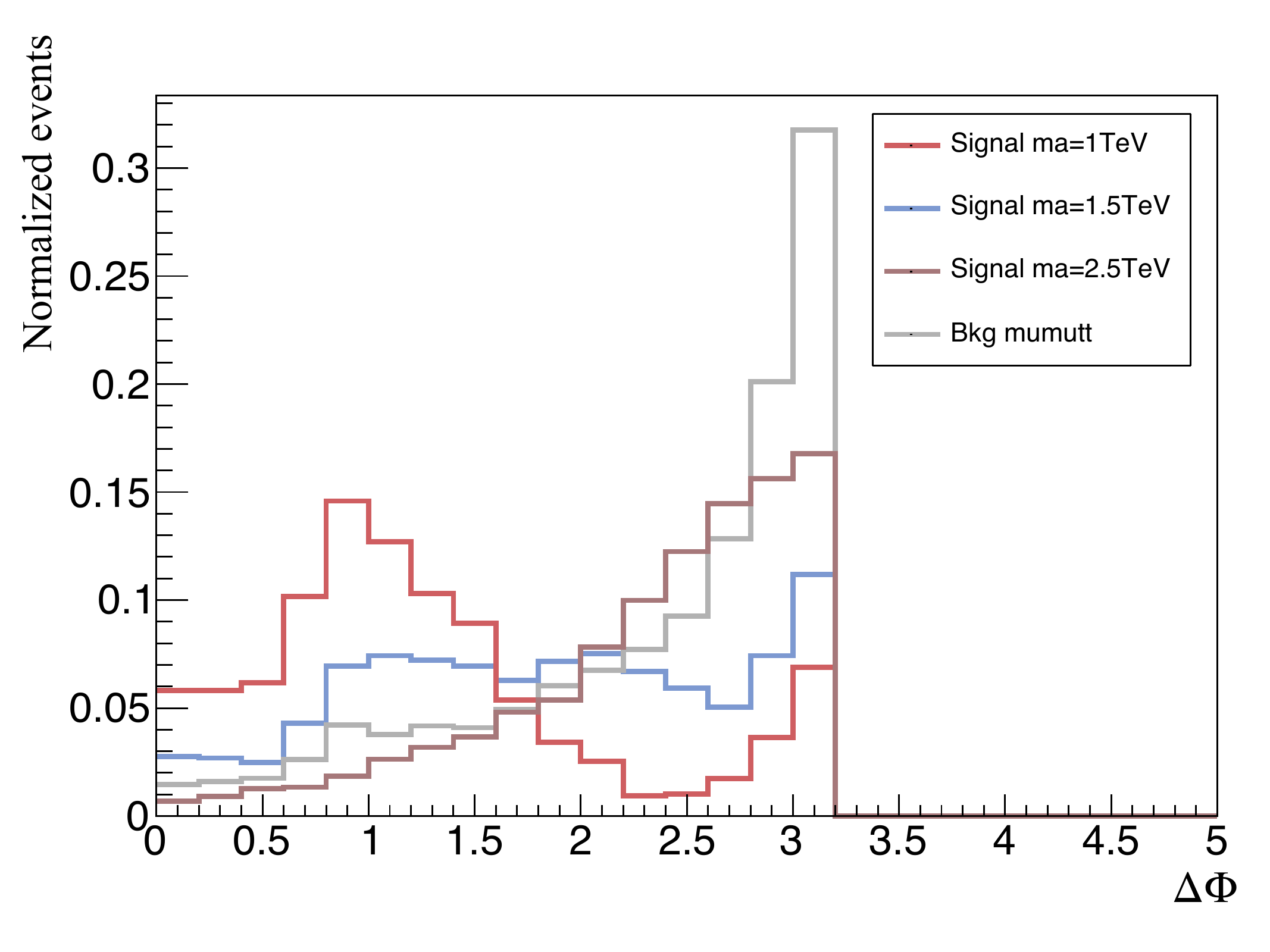}
		\subcaption{$\Delta \Phi \  ({\rm n}_{\rm fm}=1)$}
		\label{fig:Delta_Phi1}
		\end{subfigure}
\hspace{0.8cm}
    \begin{subfigure}{0.45\textwidth}
		\centering
		\includegraphics[width=\textwidth]{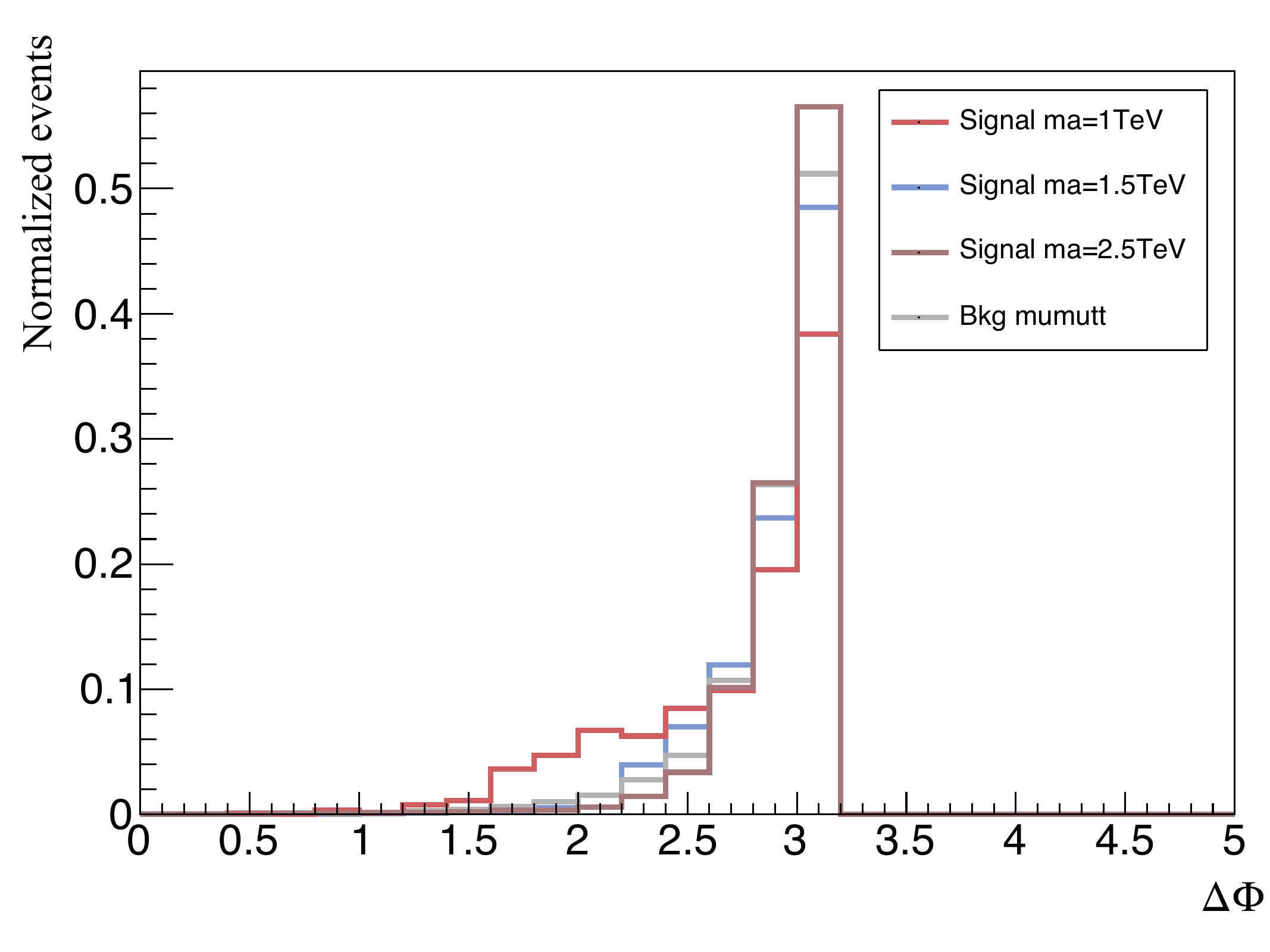}
		\subcaption{$\Delta \Phi \  ({\rm n}_{\rm fm}=2)$}
		\label{fig:Delta_Phi2}
		\end{subfigure}

	\caption{
		Distribution of signal and background events in the VT region concerning the number of FMs (a), $\mathrm{PT}_{\mathrm{sum}}$ for events with one FM $({\rm n}_{\rm fm}=1)$ (b), the azimuthal angle difference between the two top candidate jets, $\Delta \Phi$, for events with one FM $({\rm n}_{\rm fm}=1)$ (c) and two FMs $({\rm n}_{\rm fm}=2)$ (d) respectively for $\sqrt{s}=10\,\mathrm{TeV}$ and $\mathcal{L}=\SI{100}{ab^{-1}}$. 
	}
	\label{fig:n_distribution}
\end{figure}

Fig.~\ref{fig:fm} shows the distribution of the number of FMs in one event. As evident, most of the SM VBF $\mu^+ \mu^- t \bar{t}$ background events have two FMs. Although the VBF-produced ALP events mainly have zero FMs, because the $WW$ fusion process has a larger cross-section, which yields invisible neutrinos in the final state, we still focus on signal events with tagged FMs as emerging from $\gamma, Z$-induced VBFs. This is because, without the tagged FMs, one needs to consider many additional non-VBF backgrounds that we don't show in the histogram such as non-VBF $t \bar t (g)$ ($\sigma = 0.0052$ pb). Hence, we require one FM in the signal event which can exclude most of the non-VBF backgrounds and suppress the SM VBF backgrounds efficiently. We select signal events with only one FM, because if the other muon is non-forward, then the ALP which recoils against it can have a larger $p_{T}$ due to momentum conservation, which will enable us to distinguish the signal event as we will describe later.  

{Having identified the principal SM backgrounds, let us now turn to our strategy to optimize the event selection criteria and associated cut-flow analysis. For event selection, top jet candidates defined as jets with reconstructed mass $m_j$ in the interval ${140}{\ \rm GeV} < m_j < {220}{ \ \rm GeV}$ are used. Requiring two top jet candidates in the signal event helps immensely to trim down the $\mu^+ \mu^- W^+ W^-$ background, as the misidentification probability for two simultaneous $W$-jets as top-jets is quite low.} The remaining SM background $\mu^+ \mu^- t \bar{t}$ also has a pair of top jets, {and emerges as the principal background after the requirement of two top jets in the signal event.} However, {the on-shell ALP-generated top pairs possess some distinctive features}, especially on the transverse plane, which can be utilized to distinguish them.

In both the signal process $\mu^+ \mu^- (a \to t \bar{t})$ and the principal SM background $\mu^+ \mu^- t \bar{t}$, which proceed via VBF, the produced top quark pairs tend to be back-to-back in the transverse plane when there are two FMs in the event. This back-to-back nature is reflected in the distribution of the angle $\Delta \Phi$ between the transverse momenta of the top pair, defined as $\Delta \Phi \equiv \cos^{-1}\left[(\widehat{p_{T}}(t) \cdot \widehat{p_{T}} (\bar{t})\right]$ with $\widehat{p_{T}} (\cdot)$ being the direction of $t$ or $\bar{t}$ in the transverse plane; see Fig.~\ref{fig:Delta_Phi2}. However, if only one FM is present, the other outgoing muon can carry substantial transverse momentum, against which the top pair can recoil. In this case, the top quarks are no longer perfectly back-to-back, and $\Delta \Phi$ will deviate from $\pi$, depending on the transverse momentum of the non-FM. For signal events, where the top pair originates from an on-shell ALP, the transverse momentum boost of the ALP in the lab frame causes the top pair to have $\Delta \Phi$ largely deviated from $\pi$. The lighter the ALP, the more it is boosted, leading to a greater deviation of $\Delta \Phi$ from $\pi$. In contrast, for the SM background, the top pair does not necessarily come from an on-shell decay, so the two tops are boosted differently. This results in a $\Delta \Phi$ distribution that peaks around $\pi$, with a smooth tail extending away from this peak. As shown in Fig.~\ref{fig:Delta_Phi1}, in the one-FM case, a significant portion of the signal events (especially for lighter ALPs) populate the small $\Delta \Phi$ region, while the background events primarily cluster around $\pi$. In the two-FM case, since the transverse momentum taken away by the FMs is negligible, the ALP is only boosted along the z-axis. This does not affect the $\Delta \Phi$ distribution, which remains peaked around $\pi$.
 
\begin{table}[!t]\centering
	\setlength{\tabcolsep}{3mm}
    
	\begin{tabular}{|c||c|c|c|} \hline
	
		\diagbox{cut}{processes} & $\ell_\mu \bar \ell_\mu a \ (m_a=1.5{\ \rm TeV})  $ & $\ell_\mu \bar \ell_\mu t \bar{t}  $ & $\ell_\mu \bar \ell_\mu W^+ W^-  $
		\\ \hline
		origin & 199500 & 960000 & 135600000 \\ \hline
        ${\rm n}_{\rm fm} = 1$ & 44494 & 131985  & 19218587 \\ \hline
        ${\rm n}_{\rm top} = 2$ & 5819 & 3714 & \diagbox{}{} \\ \hline
        $\Delta \Phi < 2.6$ & 4415 & 1761 & \diagbox{}{} \\ \hline
        ${\rm PT}_{\rm sum} > 1400 \ {\rm GeV}$ & 3276 & 868 & \diagbox{}{} \\ \hline
	\end{tabular}

 \vspace{2mm}
  	\caption{
		Cutflow for the signal and top-rich SM processes in the VT region for $\sqrt{s}=\SI{10}{TeV}$ and $\mathcal{L}={100}{ \ \rm ab^{-1}}$. $n_{\rm fm}$, $n_{\rm top}$ are the number of FMs and number of top candidates, respectively. $\Delta \Phi$ is the angle between the momenta of the two top candidates in the transverse plane. $\text{PT}_{\rm sum}$ is the scalar sum of $p_T$ of the two top candidate jets.
	}
 \label{cutflow_table_5TeV}
\end{table}

{Similarly, we show the distribution for the transverse momentum scalar sum for the tops, denoted as $\text{PT}_{\text{sum}} \equiv \sum_{i= t, \bar t} |\vec{p}_{T} (i)|$ in Fig.~\ref{fig:PT_sum1}, where it can be seen that the heavier ALPs generate events with larger PT$_{\text{sum}}$}. The trend can be understood as follows. The ALP decay can contribute at most the mass of the ALP in the scalar sum of transverse momenta for the tops, while the recoil against the non-FM imparts some additional $|\vec{p}_T|$. For lighter ALP, the tops produced are not boosted significantly in the ALP rest frame, and the only source of $|\vec{p}_T|$ is from the recoil against the non-FM. {Although the $\Delta \Phi$ distribution does not help distinguish the signal events from the background for heavier ALPs, $\text{PT}_{\text{sum}}$ provides another distinctive filter in that region}.
\begin{figure}[t!]
    \centering
    \includegraphics[width=0.6\textwidth]{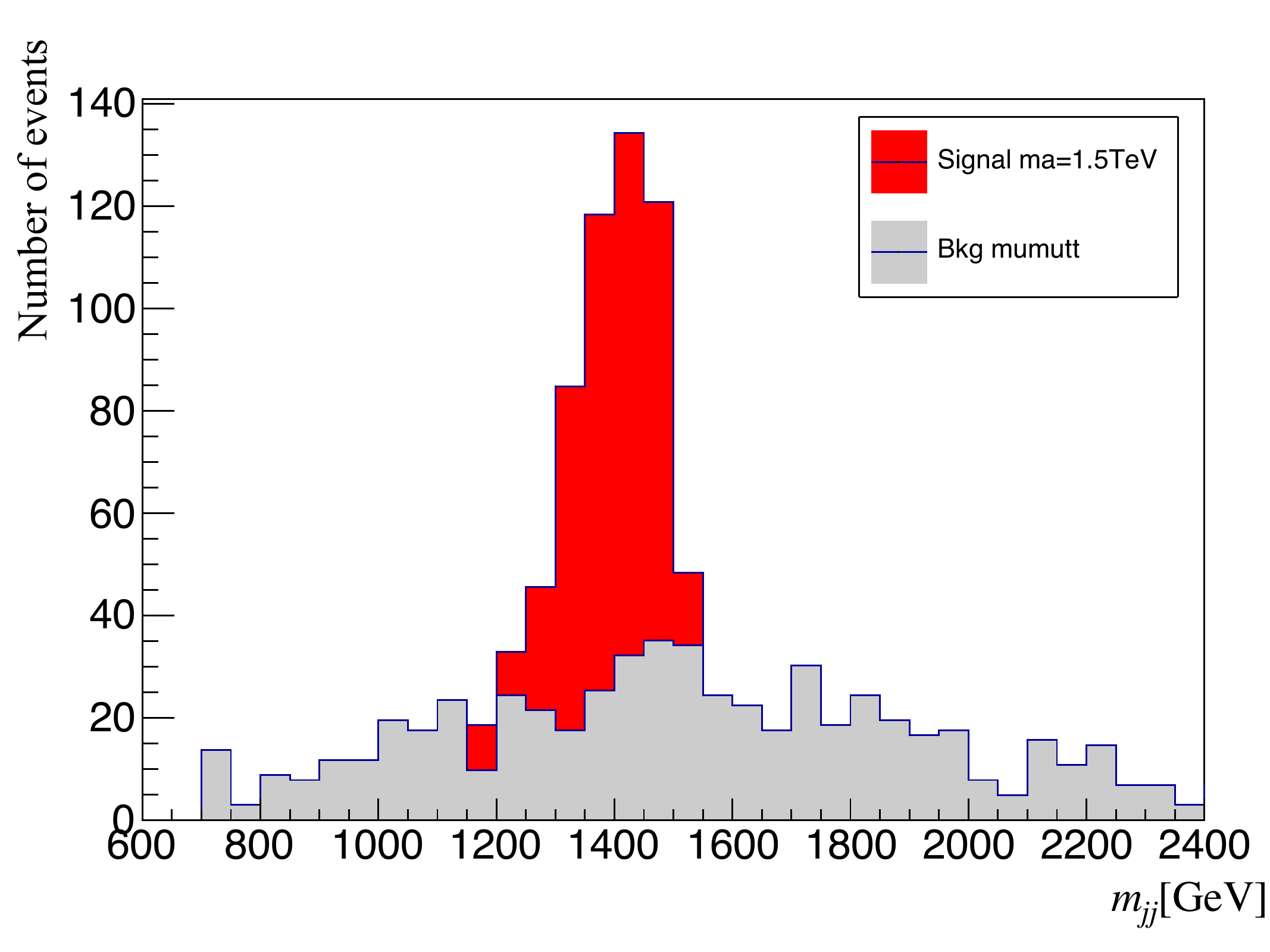}
    \caption{Dijet invariant mass distribution of the two top candidates in the VT region for $\{|C_{W}|, |C_{B}|, |C_{G}|, |C_{a \Phi}|\}/f_a=\{4, 4, 0, 6\}{\ \rm TeV}^{-1}$.}
    \label{fig:di-jet mass}
\end{figure}

 $n_{\rm fm} =1$ and $n_{\rm top} =2$ can be applied for all the events, {where $n_{\rm fm}$, $n_{\rm top}$ represent the number of FMs and top jet candidates in an event, respectively}. However, the distributions for $\Delta \Phi$ and $\text{PT}_{\text{sum}}$ vary with the ALP mass. Therefore, we use different cuts for $\Delta \Phi$, and $\text{PT}_{\text{sum}}$ according to different ALP mass ranges to maximize the selection efficiency. The signals are separated into three mass regions for the analysis, namely the {lightweight region} ($m_a = 2m_t-1.4 $ TeV), the {intermediate region} ($m_a = 1.4-2$ TeV), and the {heavyweight region} ($m_a = 2-8$ TeV), respectively. For the lightweight region, requiring $\Delta \Phi < 1.9$ is sufficient to suppress the backgrounds, while the distribution of PT$_{\rm sum}$ is similar to backgrounds. For the intermediate region, we require $\Delta \Phi < 2.6$ and PT$_{\rm sum} > 1.4$ TeV to suppress the backgrounds. For the heavyweight region, we demand PT$_{\rm sum} > 1.8$ TeV.

 In Table~\ref{cutflow_table_5TeV}, we show a cutflow for the intermediate region as an example. The background $\mu^+ \mu^- W^+ W^-$ are trimmed to a negligible amount by the $n_{\rm top}$ cut and the background $\mu^+ \mu^- t \bar t$ is suppressed efficiently by the $\Delta \Phi$ and PT$_{\rm sum}$ cuts. {Then, a clear peak around the ALP mass appears in the dijet invariant mass distribution for the top pair. We depict this for $m_a = 1.5$ TeV} in Fig.~\ref{fig:di-jet mass}, where we have scaled the couplings to $\{|C_{W}|, |C_{B}|, |C_{G}|, |C_{a \Phi}|\}/f_a=\{4, 4, 0, 6\}{ \ \rm TeV}^{-1}$ for clear depiction of the background. We will obtain the significance of the signal thus reconstructed by fitting this peak structure, which will be elaborated in the next sections.

\subsection{Statistical treatment}
\begin{figure}[t!]
    \centering
    \includegraphics[width=0.6\linewidth]{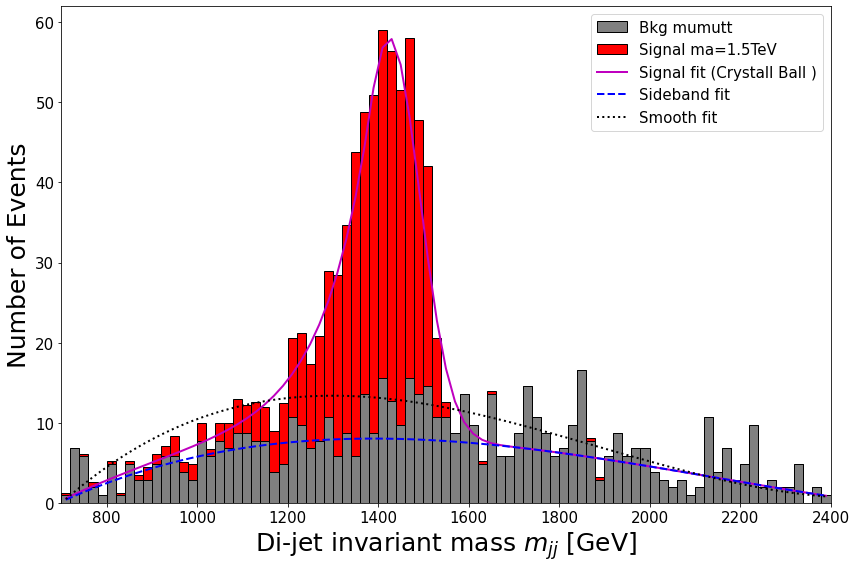}
    \caption{Result of fitting the dijet invariant mass distribution in the VT region for $\{|C_{W}|, |C_{B}|, |C_{G}|, |C_{a \Phi}|\}/f_a=\{4, 4, 0, 6\}{\ \rm TeV}^{-1}$ using Eq.~\eqref{eq:fitfunc_peak} for $m_a = 1.5$ TeV.}
    \label{fig:fitted_figs}
\end{figure}

To find the peak position and quantify the significance of the signal, we use the crystal ball function and Bernstein polynomials to fit the signal and background, respectively. {Note that the peak structure looks asymmetric and is skewed to the lower dijet invariant mass region in Fig.~\ref{fig:di-jet mass}. This is due to various lossy processes, including missing soft jets in the hadronic decay products of the two top candidates. The crystal ball function, which is effectively a skewed Gaussian distribution, is efficiently designed to model this asymmetric peak structure and can be used to recover the true ALP mass from the fit~\cite{Oreglia:1980cs, ATLAS:2023ssk}.}

Let $\lambda_i$ be the number of events in the $i^{\text{th}}$ bin of the histogram in our fitting procedure, which is decomposed as
\begin{align}
	\lambda_i = b_i + s_i
	\;.
	\label{eq:fitfunc_full}
\end{align}
Here, $b_i$ denotes the smooth distribution of background that we model with {Bernstein polynomials},
\begin{align}
	b_i = \sum_{k=0}^{3} \beta_k \ b_{k,3} (x_i)
	\;,
	\label{eq:fitfunc_bg}
\end{align}
{where $b_{k,3}$ are the Bernstein basis polynomials of degree 3, $x_i= m_{jj}^{(i)}/(m_{jj}^{\rm max}-m_{jj}^{\rm min})$ denotes the normalized dijet invariant mass with $m_{jj}^{(i)}$ being a representative value of the dijet invariant mass for the $i^{\rm th}$ bin, while the denominator is the range of $m_{jj}$ in the data,} and $\beta_k$ ($k=0 \sim 3$) are the Bernstein coefficients. On the other hand, $s_i$ denotes the peak structure of the signal contribution that we model with the crystal ball function as
\begin{align}
	s_i &= N 
    \begin{cases}
        \exp \left(-\frac{\left( m_{jj}^{(i)} - m_0\right)^2}{2\sigma^2}\right), & \ {\rm for} \ \frac{m_{jj}^{(i)} - m_0}{\sigma} > -\alpha \\
        A \left(B-\frac{m_{jj}^{(i)} - m_0}{\sigma} \right)^{-n}, & \ {\rm for} \ \frac{m_{jj}^{(i)} - m_0}{\sigma} \leq -\alpha 
    \end{cases}
	\label{eq:fitfunc_peak}
\end{align}
This crystal ball function describes an asymmetric peak structure distribution. Here we choose positive $\alpha$ and positive $n$ so that the left side of the function slowly increases and the right side decreases rather faster which models the signal well. We fit the data by $\lambda_i$ with fitting parameters being $\beta_k$ ($k=0 \sim 3$),  $N$, $A$, $B$, $m_0$, $\sigma$, $n$, and $\alpha$ for each choice of the specific $m_a$ and the coupling.
Defining a sequence of models parameterized by a parameter $\mu$ through the combination,
\begin{align}
	\tilde{\lambda}_i(\mu) = \tilde{b}_i + \mu \tilde{s}_i
	\;,
\end{align}
where $\tilde{b}_i$ and $\tilde{s}_i$ are the best-fit values of $b_i$ and $s_i$ for a given $\mu$, respectively, we treat $\tilde{\lambda}_i(\mu=1)$ to be the expected number of events for the corresponding ALP model, while $\tilde{\lambda}_i(\mu=0)$ represents the background only case. We illustrate the fitted dijet distribution with Eq.~\eqref{eq:fitfunc_peak} for $m_a=1.5$ TeV in Fig.~\ref{fig:fitted_figs}.

\begin{figure}[t!]
    \centering
    \includegraphics[width=0.65\textwidth]{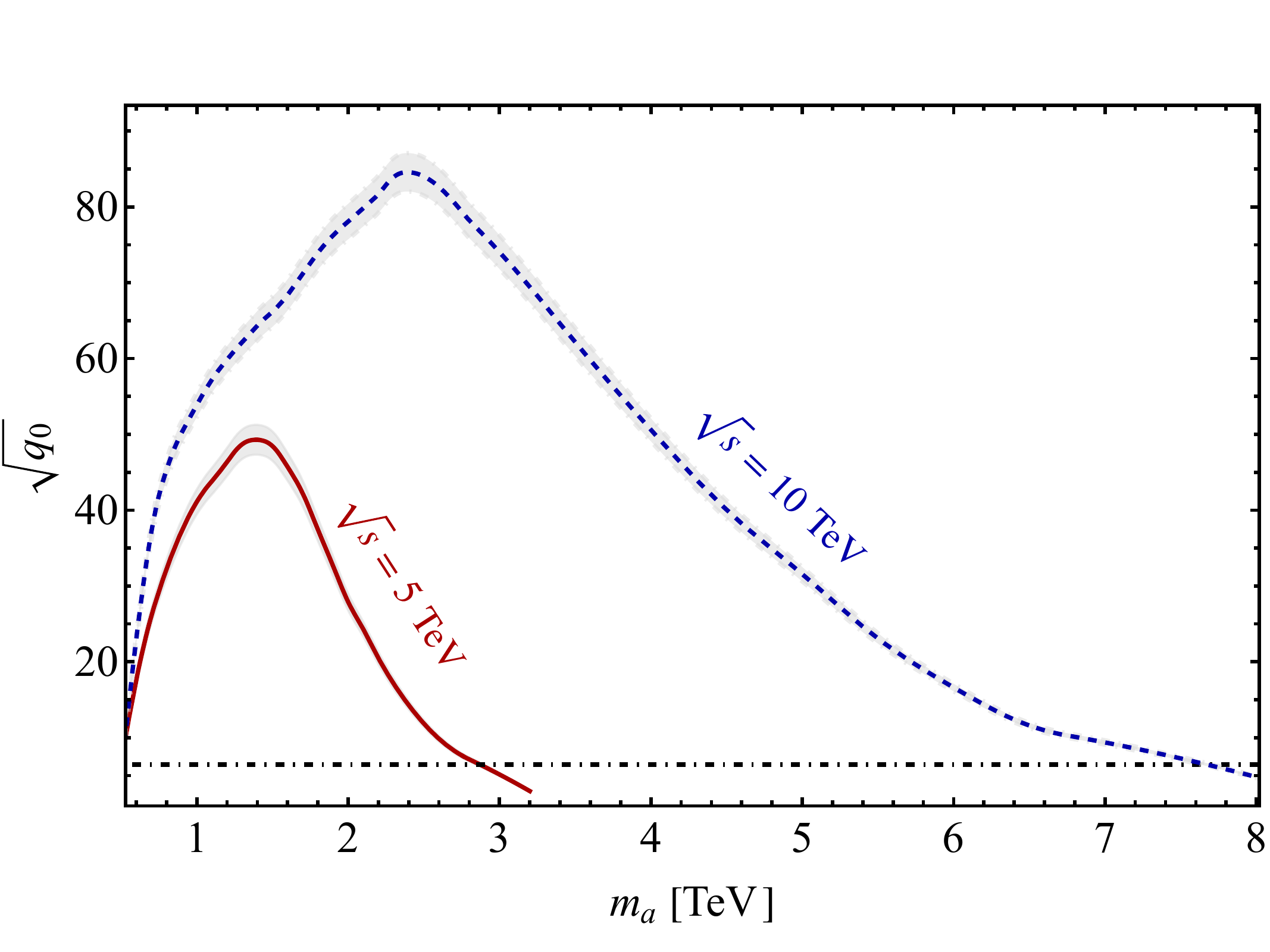}
    \caption{The test-statistic $\sqrt{q_0}$ as a function of the ALP mass $m_a$ for $\sqrt{s}=5$ TeV (red solid), and $\sqrt{s}=10$ TeV (blue dashed) with $\mathcal{L}={100}{\ {\rm ab}^{-1}}$ and  $|C_{W}|/f_a=|C_{B}|/f_a=10 {\ \rm TeV}^{-1}$. The gray band incorporates the uncertainties in the fitting procedure. The black dot-dashed line denotes $\sqrt{q_0}=6.47$ corresponding to the $5\sigma$ reach.}
    \label{fig:q_vs_ma}
\end{figure}

We use the likelihood function to calculate the significance. Let $o_i$ be the histogram obtained from the experimental data.
In our theoretical analysis, we use $o_i = \tilde{\lambda}_i (\mu=1)$ as a representative data set.
The likelihood function is defined as a function of $\mu$ as
\begin{align}
	L(\bm{o}; \mu) = \prod_{i=1}^{N_B} \frac{e^{-\tilde{\lambda}_i(\mu)}\tilde{\lambda}_i(\mu)^{o_i}}{\Gamma(o_i + 1)}
	\;,
\end{align}
with $N_B$ being the number of bins.
We also define a log-likelihood test statistic,\footnote{
In our analysis, systematic uncertainties are neglected under the assumption that the background fitting by a smooth function \cref{eq:fitfunc_bg} sufficiently reduces effects from systematic uncertainties.
}
\begin{align}
	q_0 = -2\ln \frac{L(\bm{o}; \mu=0)}{L(\bm{o}; \mu=1)}
	\label{eq:local_p}
	\;,
\end{align}
where $\mu=1$ trivially maximizes the denominator for our choice of $o_i = \tilde{\lambda}_i (\mu=1)$.
According to Wilks' theorem~\cite{10.1214/aoms/1177732360}, $q_0$ asymptotically obeys a chi-squared distribution with $7$ degrees of freedom, as inferred from Eq.~\eqref{eq:fitfunc_peak}.
Thus, we identify $\sqrt{q_0}=6.47$ as criteria for the discovery of the ALP model at the $5\sigma$ confidence level. We performed the same fitting process for different ALP masses and obtained the distribution of the significance with the ALP mass.

\subsection{Results}

\begin{figure}[t!]
    \centering
    \includegraphics[width=0.65\textwidth]{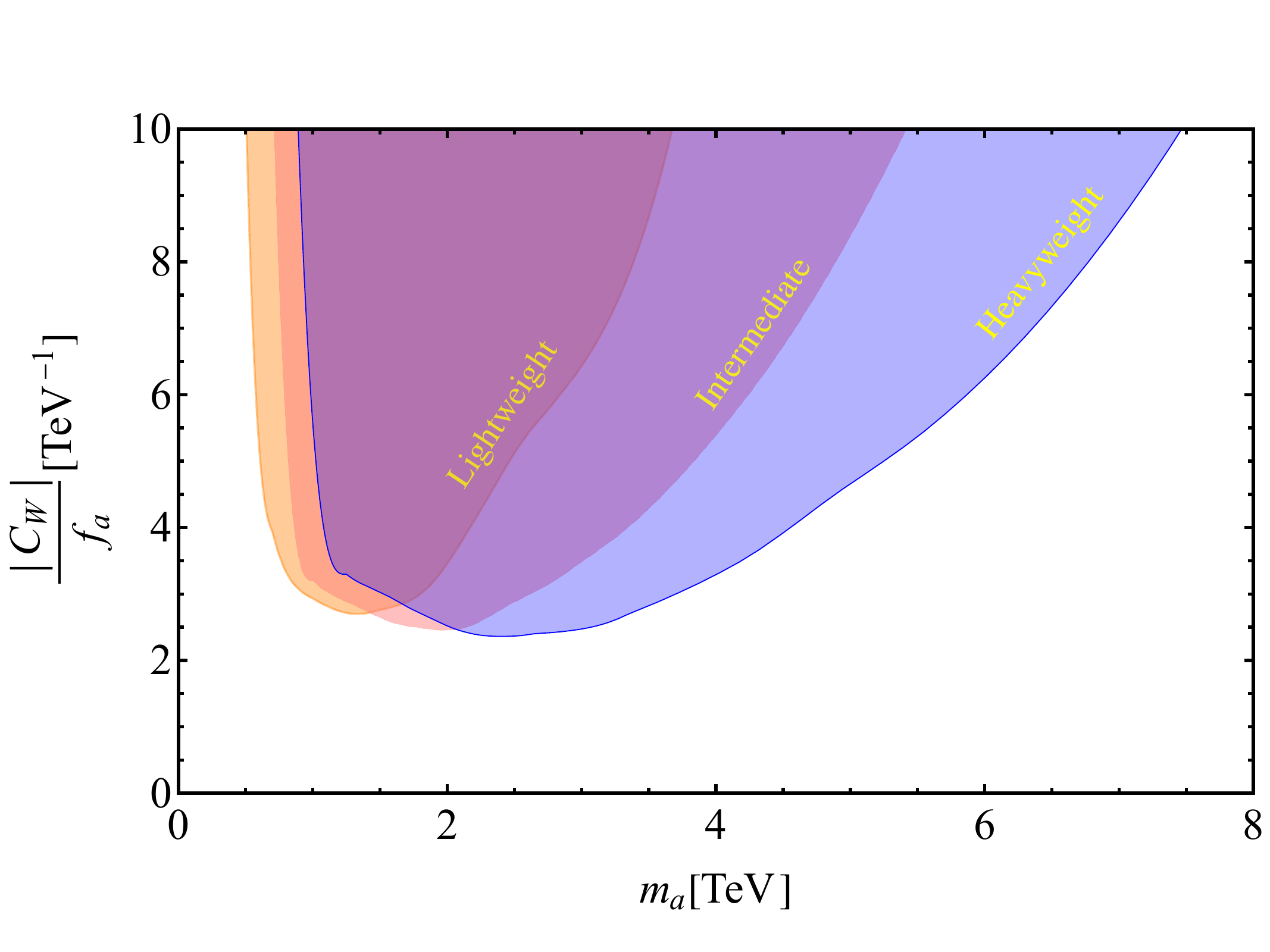}
    \caption{The $5\sigma$ contours in the normalized coupling $|C_{W}|/f_a$ and $m_a$ plane for CM energy $\sqrt{s} = 10$ TeV for integrated luminosities ${\cal L}=100$ {ab}$^{-1}$. The three contours correspond to the three cutflow methods and are the lightweight region (orange), intermediate region (pink), and heavyweight region (blue). }
    \label{fig:fa_reach_allmethod}
\end{figure}

In Fig.~\ref{fig:q_vs_ma}, we show the variation of the test-statistic $\sqrt{q_0}$ as a function of different ALP masses while keeping the coupling fixed to $|C_{W}|/f_a=|C_{B}|/f_a= 10$ TeV$^{-1}$. We assume the CM energy $\sqrt{s}=5$ TeV ($\sqrt{s}=10$ TeV) and the integrated luminosity $\mathcal{L}= 100$ ab$^{-1}$. It shows that we can probe up to $m_a \sim 3$ TeV ($8$ TeV) for $\sqrt{s}=5$ TeV (10 TeV) at 5$\sigma$.

The nature of the curves can be understood as follows: when $m_a$ is just above the threshold $2m_t$ for the $a \to t \bar{t}$ decay channel, the top quarks produced from the ALP decay have small momenta in the ALP rest frame, resulting in top decay products that are highly collimated in the lab frame. Consequently, the hadronic decay products of the top-pair are grouped as a single jet. As the ALP mass increases, the top pairs gain a larger boost, improving their detection efficiency, and explaining the upward trend of the curve. However, when the ALP mass continues to increase, the decreasing cross-section of the signal process becomes relevant, causing the curve to decline. Further, for a specific ALP mass, a higher CM energy ($\sqrt{s} = 10$ TeV) enhances the boost for the top pairs, which improves the detection efficiency of the top jets and thus enhances the significance of the ALP signals.

\begin{figure}[t!]
    \centering
    \includegraphics[width=0.65\textwidth]{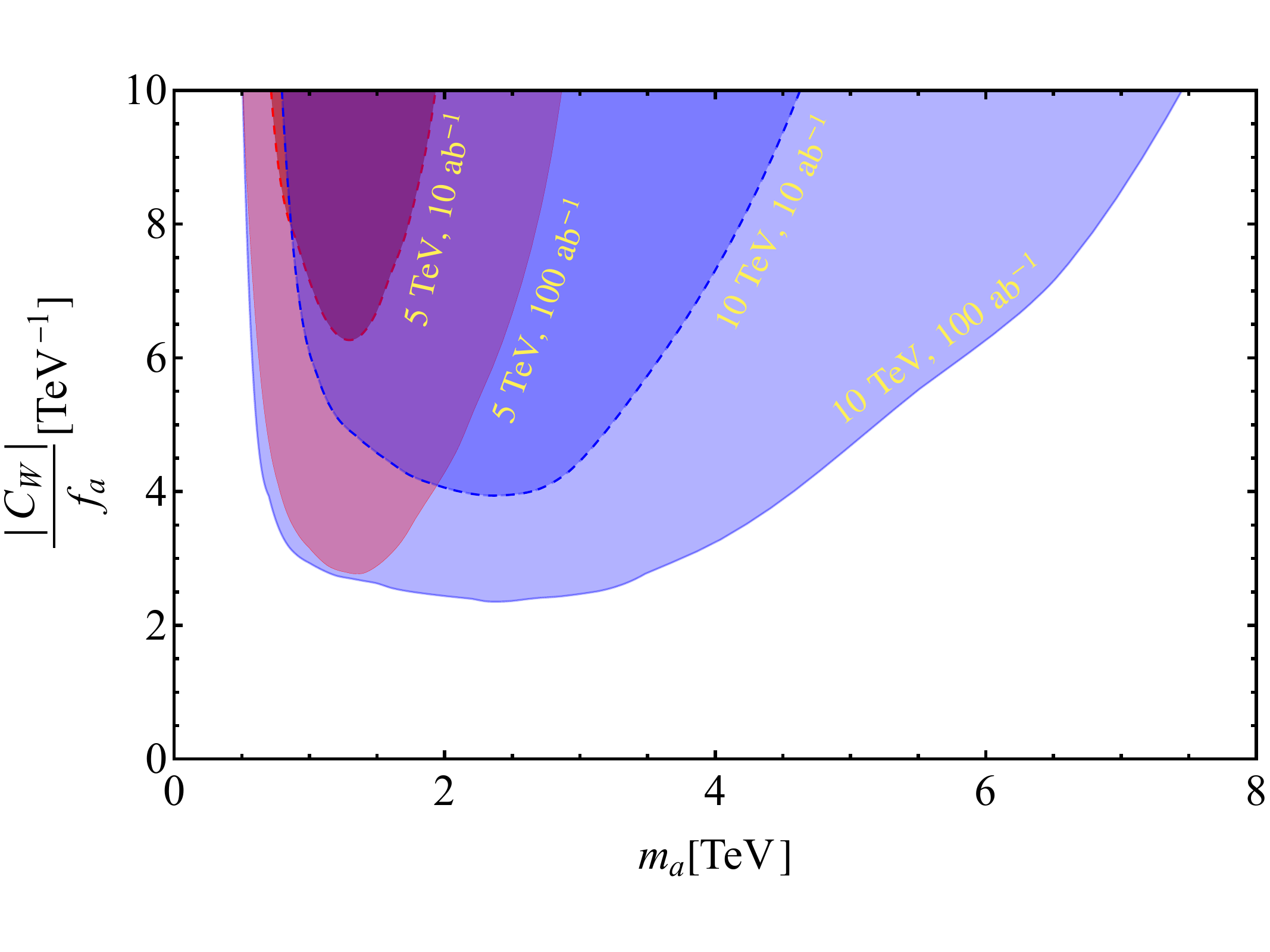}
    \caption{The $5\sigma$ reach contours in the normalized coupling $|C_{W}|/f_a$ and $m_a$ plane for two different CM energies, $\sqrt{s} = 5$ TeV (red) and $10$ TeV (blue) for integrated luminosities ${\cal L}=10$ {ab}$^{-1}$ and $100$ {ab}$^{-1}$. Note that we take $|C_{W}|/f_a$ and $|C_{B}|/f_a$ the same in the whole analysis.}
    \label{fig:VT_fa_reach}
\end{figure}

We now proceed to determine the $5\sigma$ reach in the coupling-vs-mass parameter space for the VT region. In the above analysis, we fixed the coupling values to  $|C_{W}|/f_a = |C_{B}|/f_a = 10 \, \text{TeV}^{-1}$, while the signal production cross-section scales with the square of the couplings. However, the background cross-sections are independent of the ALP couplings, meaning the test statistic does not exhibit simple scaling behavior with respect to the coupling. Nonetheless, we allow the couplings to vary and refit the dijet invariant mass distribution for different ALP masses. For each ALP mass, we calculate the minimum coupling required to achieve a significance of  $\sqrt{q_0} = 6.47$. This procedure provides the 5$\sigma$ detectable region in the  $|C_{W, B}|/f_a$-$m_a$  parameter space. Fig.~\ref{fig:fa_reach_allmethod} illustrates the 5$\sigma$ reach for three distinct cutflows, as described earlier, with $\sqrt{s} = 10 \, \text{TeV}$, and $\mathcal{L} = 100 \, \text{ab}^{-1}$. The advantage of using the three specialized cut flows for different ALP mass regions is visible. Finally, we combine these results to obtain the cumulative reach as shown in Fig.~\ref{fig:VT_fa_reach}, where we also compare the case for  $\sqrt{s} = 5 \, \text{TeV}$.

\section{Region TT}
\label{sec:TT}

{The region TT, where the ALP is produced by the top-associated production and decays promptly to $t \bar t$ was analyzed in our previous work~\cite{Chigusa:2023rrz}. The process under consideration is $\mu^+ \mu^- \to t \bar t (a \to t \bar t)$, and we look for signal events with four top jets in the final state. Here we will revisit this region and demonstrate that a choice of a larger $R$ parameter for the top jet reconstruction enhances the signal selection efficiency and improves the reach in the parameter space in the TT region. Let us summarize the event selection cutflow and principal backgrounds for the TT region here.} SM processes with top-jet-rich final states can serve as backgrounds in our analysis. The dominant background arises from SM four-top production ($t \bar{t} t \bar{t}$), which shares the same event topology as the signal. Subdominant backgrounds include processes where a top-quark pair is produced alongside other heavy jets that can be misidentified as top jets, such as $t \bar{t} W^+ W^-$, $t \bar{t} h$, and $t \bar{t} Z$. To enhance signal sensitivity, we focus on events containing at least three top candidates. This selection efficiently suppresses the backgrounds from $t \bar{t} W^+ W^-$, $t \bar{t} h$, and $t \bar{t} Z$, while retaining most of the top pairs originating from ALP decays. Additionally, since signal events tend to produce a higher jet multiplicity, we require more than three jets per event. After applying these selection criteria, a peak structure emerges around the ALP mass in the di-jet invariant mass distribution. To extract the signal significance, we model the background using Bernstein basis polynomials and fit the signal with a Gaussian distribution function.

\begin{table}[!t]\centering
	\setlength{\tabcolsep}{3mm}
	\begin{tabular}{|c||c|c||c|c|} \hline
	
		\diagbox{cut}{processes} & $ t \bar{t} a$ (R=0.5) & Bkg (R=0.5) & $ t \bar{t} a$ (R=0.8) & Bkg (R=0.8)   
		\\ \hline
		origin & 9015 & 96965 & 9015 & 96965  \\ \hline
        $n_{\rm top}\geq 3$ & 495 & 148 & 619 & 217  \\ \hline
        $n_{\rm jet}  \geq 4$ & 417 & 84 & \diagbox{}{}  & \diagbox{}{} \\ \hline
	\end{tabular}
 \vspace{2mm}
  	\caption{
		Cutflow for the signal process in the TT region with benchmark parameters, $|C_{a \Phi}|/f_a = 6{ \ \rm TeV}^{-1}$ and $m_a = 1{ \ \rm TeV}$ and Bkg denotes cumulative top-rich SM processes ($t \bar t t \bar t$, $t \bar t W^+ W^-$, $t \bar t h$ and $t \bar t Z$) for $\sqrt{s}=5 {\rm \ TeV}$ and $\mathcal{L}={100}{ \ \rm ab^{-1}}$.
	}
 \label{TT_cutflow_table_5TeV}
\end{table}

Identification of top jets plays an important role in this analysis because one relies on the three top candidate events. A large $R$ parameter helps enhance acceptance of the boosted jets by collecting top decay products even for a mild boost. On the other hand, if $R$ is too large, the mis-identification rate of the top jet candidates increases due to the several energetic partons clustered into a single jet. Thus, the choice of $R$ should be optimized to achieve the optimal sensitivity for the ALP.

\begin{figure}[t!]
    \centering
    \includegraphics[width=0.65\textwidth]{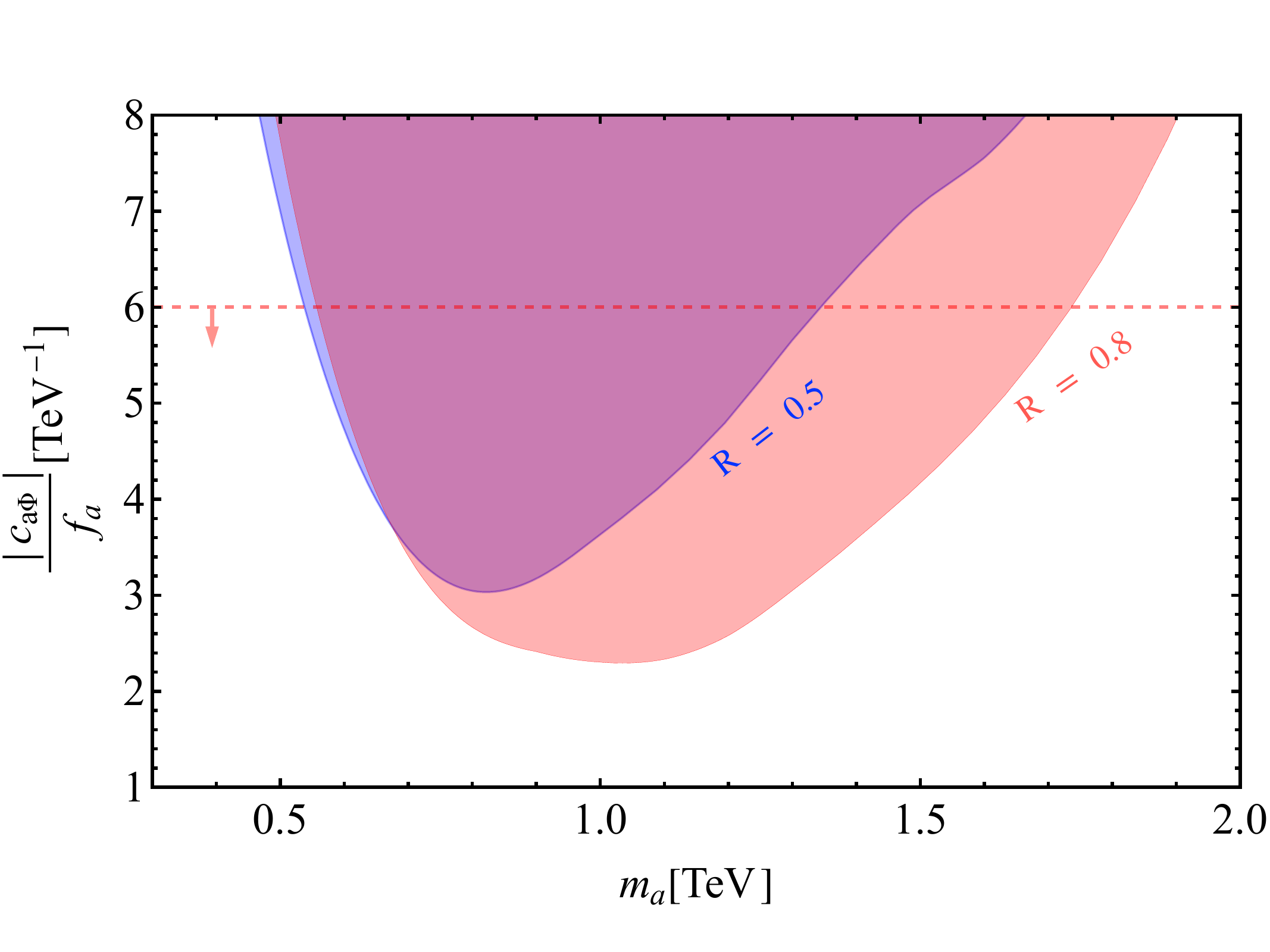}
    \caption{The $5\sigma$ reach contours in the normalized coupling $|C_{a \Phi}|/f_a$ and $m_a$ plane for CM energy $\sqrt{s} = 5$ TeV for integrated luminosities $100$ {ab}$^{-1}$. The two contours denote the choices $R=0.5$ (blue) and $R=0.8$ (red). The approximate unitarity bound is shown as the red dashed line~\cite{Brivio:2021fog}.}
    \label{fig:TT_fa_reach}
\end{figure}

We use the same benchmark as the previous work, $|C_{a \Phi}|/f_a = 6 \ {\rm TeV}^{-1}$ with turning off other couplings and take $m_a = 1 \ {\rm TeV}$ for CM energy $\sqrt{s}= 5$ TeV and $\mathcal{L}=100 \ {\rm ab^{-1}}$. We tried several choices for $R$, and $R=0.8$ is found to optimize the top identification efficiency. Table~\ref{TT_cutflow_table_5TeV} shows a comparison of the cutflow where the Bkg represents the sum of SM backgrounds, $t \bar t t \bar t$, $t \bar t W^+ W^-$, $t \bar t h$ and $t \bar t Z$. Note that for larger $R$, $n_{\rm jet} \geq 4$ is not required because the total number of jets is decreased. For the $R=0.8$ case, the signal events with 3 top candidates increase significantly. Although the backgrounds also increase, the distribution of dijet invariant mass is smooth so the additional backgrounds do not deteriorate the signal significance. Following the same analysis strategy, we fit the data for different ALP masses and revise the parameter space reach. We show the final result for $R=0.5$ {(same as our previous work in Ref.~\cite{Chigusa:2023rrz})} and $R=0.8$ in Fig.~\ref{fig:TT_fa_reach}. With a larger $R$ parameter, we can detect heavier ALP to about 1.7 TeV at 5$\sigma$. We note that increasing the top identification efficiency with a large $R$ parameter is beneficial for small CM energy. When the CM energy becomes larger, top jets are more boosted and have a smaller radius so that $R=0.5$ is desirable for $\sqrt{s}=10 \ {\rm TeV}$ case.

\section{Region VG and TG}
\label{sec:VGTG}

In the VG region, the dominant ALP production is through the VBF process, while the ALP decays to $gg$ primarily. Due to the similar VBF production process of ALP as in the VT region, the FMs and PT$_{\rm sum}$ cuts are also effective in suppressing the background events. However, unlike the top jets from ALP decay in the VT region, which have a unique feature of heavy jet mass, it is challenging to distinguish the gluon jets from other light quark-initiated jets. By requiring one FM, the non-VBF backgrounds are suppressed. The remnant VBF backgrounds are mainly $\mu^+ \mu^- W^+W^-$ and $\mu^+ \mu^- q\bar{q}$, {where $q$ denotes any light SM quark}. The cross-sections of these backgrounds are significantly large compared to the signal. They are still several orders of magnitude larger than the signal, even if we choose the coupling as $\{|C_{W}|, |C_{B}|, |C_{G}|, |C_{a \Phi}|\}/f_a=\{10, 10, 3, 0\} \ {\rm TeV}^{-1}$. With these parameter settings and $m_a=1 \ {\rm TeV}$, $\sqrt{s}=\SI{10} \ {\rm TeV}$, $\mathcal{L}=\SI{100} \ {\rm ab^{-1}}$, we simulate the events to illustrate it. 


\begin{table*}[!t]\centering
	\setlength{\tabcolsep}{3mm}
	\begin{tabular}{|c||c|c|c|c|} \hline
		&$ \ell_\mu \bar \ell_\mu a \left(m_a=1{\rm TeV}\right)$ &  $ \ell_\mu \bar \ell_\mu W^+W^-$  & $ \ell_\mu \bar \ell_\mu q\bar{q}$ 
		\\ \hline
		$\sigma$[pb] & 0.0018 & 1.356 & 1.849 \\ \hline
		
	\end{tabular}
 \vspace{0.1cm}
	\caption{
	Cross-sections for the signal and dominant backgrounds for the VG region for $\sqrt{s}=10\, \ \mathrm{TeV}$, $\{|C_{W}|, |C_{B}|, |C_{G}|, |C_{a \Phi}|\}/f_a=\{10, 10, 3, 0\}{ \ \rm TeV}^{-1}$, and $m_a = 1 {\ \rm TeV}$. }
 \label{tab:VG_cross_sections}
\end{table*}

\begin{figure}[!t]
	\centering
		\begin{subfigure}{0.45\textwidth}
		\centering
		\includegraphics[width=\textwidth]{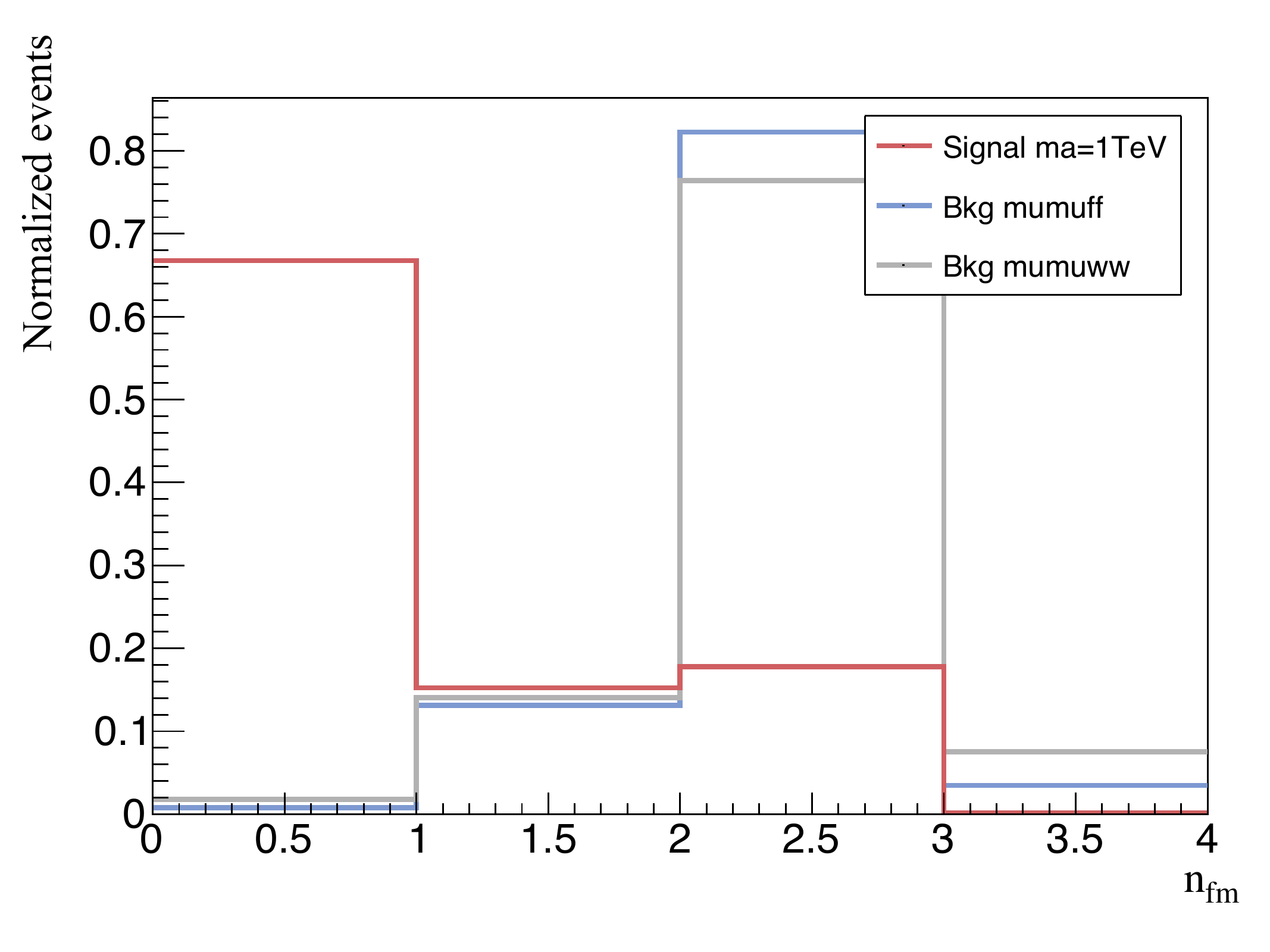}
		\subcaption{Number of FMs}
		\label{fig:vg_fm}
		\end{subfigure}   
\hspace{0.8cm}
		\begin{subfigure}{0.45\textwidth}
		\centering
		\includegraphics[width=\textwidth]{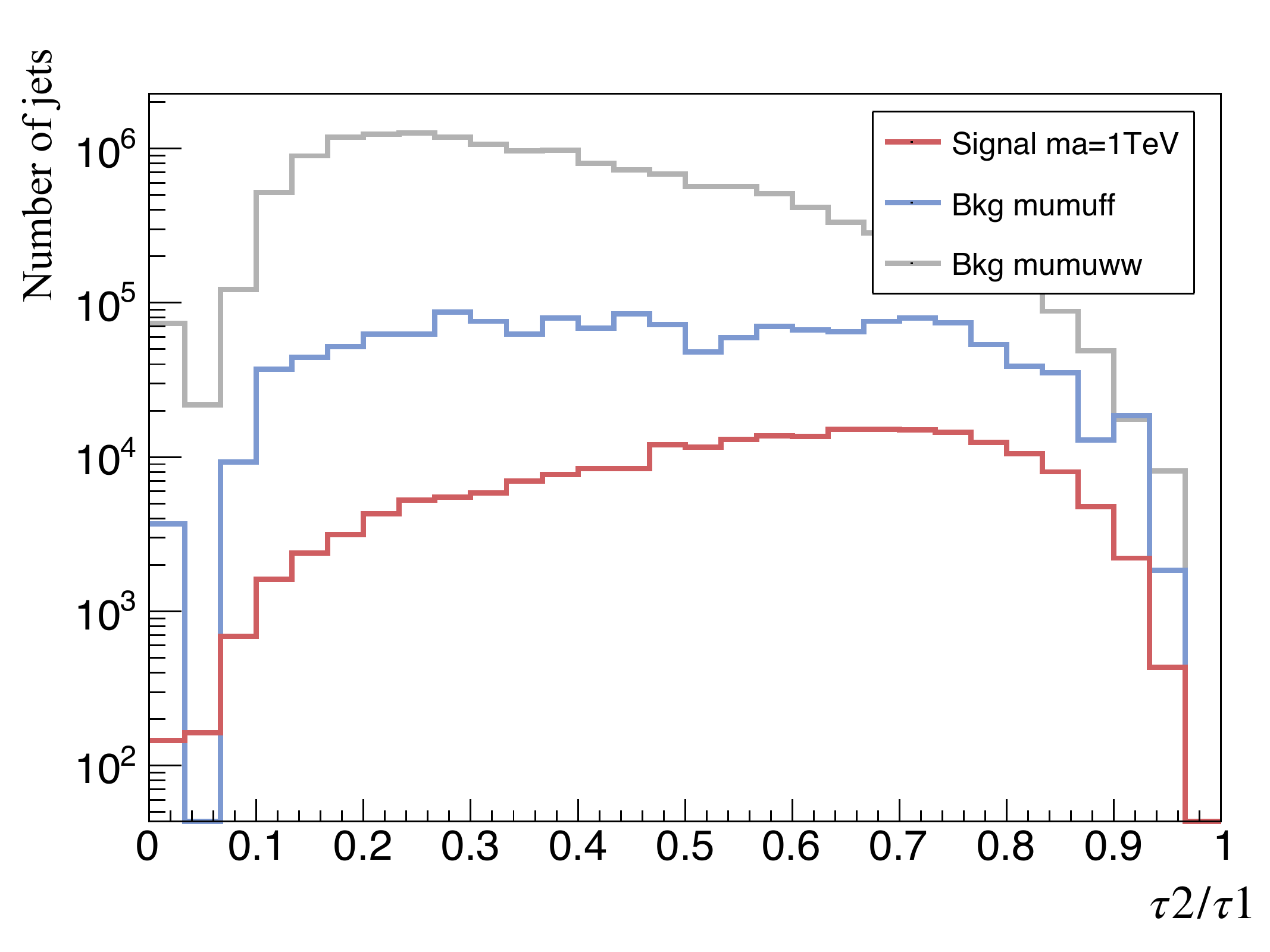}
		\subcaption{$\tau_2/\tau_1$}
		\label{fig:vg_tau}
		\end{subfigure}

	\caption{
		(a) Distribution of signal and background events for the number of FMs and (b) $\tau_2/\tau_1$ of jets for events with one FM for $\sqrt{s}=10\,\ \mathrm{TeV}$ and $\mathcal{L}={100}{ \ \rm ab^{-1}}$.
	}
	\label{fig:tau_distribution}
\end{figure}

In addition to the cut of the number of FMs and PT$_{\rm sum}$, a cut based on $N$-subjettiness is considered to suppress the $\ell_\mu \bar \ell_\mu W^+W^-$~\cite{Thaler:2010tr}. {$N$-subjettiness, denoted as $\tau_N$, is a metric that evaluates how effectively a jet can be characterized as having $N$ subjets, corresponding to $N$ energetic partons. One reconstructs a candidate jet and identifies $N$ candidate subjets using the exclusive $k_T$ clustering algorithm. With these candidate subjets, $\tau_N$ is calculated via} 
\begin{align}
	\tau_N = \frac{1}{d_0} \sum_k p_{T,k} {\rm min} \{\Delta R_{1,k}, \Delta R_{2,k}, \ ... \ , \ \Delta R_{N,k}, \}
	\label{eq:N_subjettiness}
	\;,
\end{align}
{where $k$ runs over the constituent particles in a given jet and $p_{T, k}$ are their transverse momenta.
$\Delta R_{J,k}$ = $\sqrt{(\Delta \eta)_{J,k}^2+(\Delta \phi)_{J,k}^2}$ is the distance in the rapidity-azimuth plane between a candidate subjet $J$ and a constituent particle $k$. $d_0$ is the normalization factor and is defined as}
\begin{align}
	d_0 =  \sum_k p_{T,k} R_0 
	\label{eq:d0norm}
	\;,
\end{align}
{where $R_0$ is the characteristic jet radius used in the jet clustering algorithm. If all the radiation of the jets is aligned with the candidate subjet directions, the $\tau_N$ of the jets is close to zero and the jets have $N$ (or few) subjets. If a large fraction of the jet's energy is distributed away from the candidate subjet directions, the $\tau_N$ of the jets is significantly larger than zero and they have at least $N+1$ subjets.

Gluon jets tend to have more complicated sub-structures than the $W$-jets, therefore, a large value of $\tau_2/\tau_1$ helps suppress SM backgrounds associated with $W$-jets}~\cite{Thaler:2010tr}. We require that $\tau_2/ \tau_1$ of {at least two jets in an event} should be larger than $0.6$ to suppress $\ell_\mu \bar \ell_\mu W^+W^-$ according to Fig.~\ref{fig:vg_tau}. Note, however, that more involved attempts to veto $W$-jets are not useful because of the low $W$ reconstruction efficiency due to the insufficient boost of $W$ bosons. Table~\ref{tab:VG_cross_sections} shows the cross-section of signal and backgrounds. To suppress the background, one FM and a large PT$_{\rm sum}$ are required in addition to the cut based on $\tau_2/\tau_1$.
The cutflow is shown in Table~\ref{TG_cutflow_table}. Naively, a simple cut-and-count gives the signal significance $S/\sqrt{B} \sim 25$ with $S\sim 5\times 10^3$ and $B\sim 4\times 10^4$ being the number of signal and background events after the cut. However, since it is difficult to characterize the signal by, \textit{e.g.}, fitting the peak of the dijet invariant mass as in the case of regions VT and TT, this analysis is highly vulnerable to systematic errors such as the fluctuation of integrated luminosities. Assuming $\Delta \mathcal{L}/\mathcal{L} = 5\,\%$ of fluctuation, the signal significance now reduces to
\begin{align}
	\frac{S}{\sqrt{B + B^2 (\Delta \mathcal{L}/\mathcal{L})^2}} \sim 2.5,
\end{align}
which is well below the discovery criteria.

We conclude that in the VG region, the overwhelming background cross-section and the lack of distinctive signal features make it challenging to discover the ALP signal with a sufficiently high significance level. Without a reliable method to distinguish gluon jets from other types of jets, the signal remains difficult to isolate and reconstruct. However, if future advancements in jet substructure techniques enable the separation of gluon jets from light quark jets and $W$-jets, this region may become accessible~\cite{Gallicchio:2012ez, Nakai:2020kuu, Karagiorgi:2021ngt, Bonilla:2022wzp}.

Finally, let us comment on the TG region, where the ALP is produced in association with a top quark and subsequently decays into a pair of gluons, \textit{i.e.,} the process under consideration is $\mu^+ \mu^- \to t \bar t (a \to gg)$. Events containing two top candidates and gluon jets have the potential to be signal events, with the ALP mass extracted by fitting the invariant dijet mass distribution. However, several challenges arise in this region. First, the signal cross-section is relatively small, approximately $2.7 \times 10^{-6}$ pb for the benchmark scenario $\{|C_{W}|, |C_{B}|, |C_{G}|, |C_{a \Phi}|\}/f_a=\{1, 1, 3, 1\} \ {\rm TeV}^{-1}$ at $\sqrt{s} = 5 \  {\rm TeV}$. In contrast, background processes such as  $t \bar t q \bar q$  ($\sigma = 1.4 \times 10^{-3}$ pb),  $t \bar t W^+ W^-$  ($\sigma = 4.7 \times 10^{-4}$ pb),  $t \bar t h$  ($\sigma = 3.2 \times 10^{-4}$ pb), and  $t \bar t Z$  ($\sigma = 1.6 \times 10^{-4}$ pb), where  $q$  and  $\bar{q}$  represent SM quarks and anti-quarks, respectively, contribute significantly and poses significant obstacle. Furthermore, due to the absence of unique discriminating characteristics such as FMs or differentiating distribution of $\text{PT}_{\rm sum}$, background suppression remains inefficient. This makes the TG region even more challenging than the VG region.

\begin{table}[!t]\centering
	\setlength{\tabcolsep}{3mm}
    
	\begin{tabular}{|c||c|c|c|} \hline
	
		\diagbox{cut}{processes} & $\ell_\mu \bar \ell_\mu a \ (m_a=1{\rm TeV})  $ & $\ell_\mu \bar \ell_\mu W^+W^-$ & $\ell_\mu \bar \ell_\mu q\bar{q}$
		\\ \hline
		origin & 181400 & 135600000  & 184900000 \\ \hline
        ${\rm n}_{\rm fm} = 1$ & 27645 & 18759099  & 24274766 \\ \hline
        $\tau_2/\tau_1 > 0.6$ & 13151 & 3137784 & 863483 \\ \hline
        ${\rm PT}_{\rm sum} > 1000$ GeV & 5169 & 27120 & 14792 \\ \hline
	\end{tabular}

 \vspace{2mm}
  	\caption{
		Cutflow for the signal process and SM VBF processes in VG region for $\sqrt{s}=\SI{10}{TeV}$ and $\mathcal{L}={100}{ \ \rm ab^{-1}}$. $n_{\rm fm}$ is the number of FMs. ${\rm PT}_{\rm sum}$ is the scalar sum of $p_{\rm T}$ of the top pairs.
	}
 \label{TG_cutflow_table}
\end{table}

\section{Discussions}
\label{conclusion}

The clean, high-energy environment of a future muon collider with CM energy in the range of $\mathcal{O}(1-10)$ TeV provides a promising setting for searching for TeV-scale ALPs with TeV-scale decay constants. In this study, we explored potential detection channels at a future muon collider, analyzing the interplay among different ALP couplings from an effective theory perspective. We categorized the search regions into five types—TT, TG, VT, VG, and VV—based on ALP production mechanisms, either via VBF or top-associated production, and their respective decay channels. We identified each region in the parameter space and summarized it in Fig.~\ref{fig:ALP_production_decay}. Apart from the VV region, which has been extensively studied in the literature, we found that the VT region can be probed across a wide mass range with the help of forward muon detectors. The reach in the TT region can be further enhanced by optimizing the cone size parameter, improving upon our previous study~\cite{Chigusa:2023rrz}. While the VG and TG regions pose significant challenges associated with background contamination, the VG region remains promising with improved machine learning techniques to distinguish gluon jets from light quark jets and $W$-jets~\cite{Gallicchio:2012ez, Nakai:2020kuu, Karagiorgi:2021ngt, Bonilla:2022wzp}. 

While we have employed an effective field theory (EFT) framework to analyze ALP phenomenology in a model-independent manner, caution is required when translating future constraints into the parameter space of specific ALP models, particularly in regions where the momentum transfer may exceed the effective cut-off scale $\sim 4 \pi f_a$. To ensure the validity of our EFT approach, we have presented unitarity limits within our projected parameter space reach, indicating that our analysis remains reliable where the couplings lie below these limits. Nevertheless, since VBF production is enhanced in the soft and collinear limit, the EFT treatment is expected to remain valid, as the momentum transfer at the ALP production vertex stays below the cut-off. For a future study, it would be interesting to explore specific UV completions and examine how the reach of the ALP parameter space deviates from the EFT expectations.

\section*{Acknowledgements}

SC is supported by the Simons Foundation.
SC is supported by the U.S. Department of Energy, Office of Science, the BNL C2QA award under grant Contract Number DE-SC0012704 (SUBK\# 390034).
YN is supported by Natural Science Foundation of Shanghai.

\appendix

\bibliographystyle{jhep}
\bibliography{rsc_bib}

\end{document}